\documentclass[preprint]{aastex631}
\newcommand{\gtsim}{\raisebox{-1.0ex}{$\stackrel{\textstyle>}{\sim}$}}
\newcommand{\ltsim}{\raisebox{-1.0ex}{$\stackrel{\textstyle<}{\sim}$}}
\long\def\comment#1{}
\def\goes{{\sl GOES}}
\def\yohkoh{{\sl Yohkoh}}

\def\hinode{{\sl Hinode}}

\def\p78{{\sl P78-1}}

\def\sdo{{\sl SDO}}
\def\iris{{\sl IRIS}}

\def\fexxiv{Fe~{\sc xxiv}}

\def\feix{Fe~{\sc ix}}
\def\fexii{Fe~{\sc xii}}

\def\fexviii{Fe~{\sc xviii}}

\def\heii{He~{\sc ii}}

\def\civ{C~{\sc iv}}

\def\etal{et~al.}

\begin{document}
%

\title{Solar Active Region Coronal Jets. III. Hidden-Onset Jets}

\author{Alphonse C. Sterling}
\affiliation{NASA/Marshall Space Flight Center, Huntsville, AL 35812, USA}

\author{Ronald L. Moore} 
\affiliation{Center for Space Plasma and Aeronomic Research, \\
University of Alabama in Huntsville, Huntsville, AL 35805, USA}
\affiliation{NASA/Marshall Space Flight Center, Huntsville, AL 35812, USA}

\author{Navdeep K. Panesar} 
\affiliation{Bay Area Environmental Research Institute, NASA Research Park, Moffett Field, CA 94035, USA}
\affiliation{Lockheed Martin Solar and Astrophysics Laboratory, 3251 Hanover Street, Building 252, Palo Alto, CA 94304, USA}

\comment{

\author{\etal}

\author{Louise K. Harra} 
\affiliation{Physikalisch Meteorologisches Observatorium Davos, World Radiation Center, 7260 Davos, Switzerland}
\affiliation{Institute for Particle Physics and Astrophysics, ETH Z{\"u}rich, 8092 Z{\"u}rich, Switzerland}

} 

\begin{abstract}

Solar quiet- and coronal-hole region coronal jets frequently clearly originate from erupting minifilaments, but active-region 
jets often lack an obvious erupting-minifilament source.  We observe a coronal-jet-productive active region 
(AR), AR~12824, over 2021 May~22 0---8\,UT, primarily using Solar Dynamics Observatory (\sdo) Atmospheric Imaging Array (AIA) EUV images and \sdo/Helioseismic and Magnetic Imager (HMI) magnetograms.  
Jets were concentrated in two locations in the AR: on the south side and on the northwest side of the AR's lone 
large sunspot. The south-location jets are 
oriented so that we have a clear view of the jets' origin low in the atmosphere: their source is clearly minifilaments erupting from
locations showing magnetic flux changes/cancelations.  After erupting a projected distance $\ltsim$5$''$ away from their 
origin site, the minifilaments erupt outward onto far-reaching field as part of the jet's spire, quickly losing their minifilament character.  
In contrast, the northwest-location jets show no clear erupting minifilament, but the source site of those jets are obscured 
along our line-of-sight by absorbing chromospheric material.  EUV and magnetic data indicate that the likely source 
sites were $\gtsim$15$''$ from where the we first see the jet spire; thus an erupting minifilament would likely lose its minifilament
character before we first see the spire.  We conclude that such AR jets could work like non-AR jets, but the 
erupting-minifilament jet source is often hidden
by obscuring material.  Another factor is that magnetic eruptions making some AR jets carry only a harder-to-detect comparatively 
thin ($\sim$1---2$''$) minifilament ``strand."

\end{abstract}

\keywords{Solar filament eruptions, solar extreme ultraviolet emission, solar active regions, solar active region magnetic fields}

\section{Introduction}
\label{sec-introduction}

Solar coronal jets appear as outflows of coronal plasma, usually seen in soft X-ray (SXR) or EUV images,
that extend to be long and narrow spires emanating from a bright base region \citep{shibata.et92}.  They 
are common in all regions of the Sun, including coronal holes \citep{cirtain.et07}, quiet Sun, and the periphery of
active regions \citep[e.g.,][]{shimojo.et96}. 

Studies with \yohkoh's soft X-ray telescope, \hinode's X-ray Telescope (XRT), and magnetographs on the
ground and in space have contributed to our understanding of coronal jets, and several summary works describing 
them are now available \citep[e.g.,][]{shibata.et11,raouafi.et16,hinode.et19,shen21,schmieder22}.  Especially with the benefit 
of images from the Atmospheric Imaging Assembly (AIA) on the Solar Dynamics Observatory (\sdo), which provides 
on a regular and consistent basis imaging of the entire Sun in EUV at time cadences and spatial resolution superior to 
instruments that preceded it, it has been possible to put together a scenario that describes the origin of many, if not most 
or all, coronal jets in coronal holes and quiet-Sun regions. They are the consequence of solar eruptions that are similar
to the solar eruptions that make typical solar flares and CMEs, but occurring on a size and energy scale substantially 
smaller than those better-resolved larger events.  They form at the locations of small-scale eruptions that often eject 
a minifilament \citep[e.g.,][]{shen.et12,adams.et14,sterling.et15}, and produces a miniature flare at the eruption location 
\citep{sterling.et15}.  In coronal holes 
and quiet Sun the erupting minifilaments have sizes of $\sim$10$^4$\,km \citep{sterling.et15,sterling.et22,panesar.et16a},
which is about a factor of three to more-than-ten smaller than filaments that erupt to make typical flares and 
CMEs \citep{bernasconi.et05}.  The brightening at the location of the eruption is analogous to the
typical flares beneath typically sized erupting filaments.  In the jet-producing case, we call the brightening a ``jet bright point," or JBP,
following \citet{sterling.et15}.

The magnetic setup of the minifilament-eruption region where jets form is that of a magnetic anemone region
\citep{shibata.et07}, with a minority-polarity flux patch embedded in surrounding majority-polarity flux, and a magnetic null
point in the corona elevated above the majority-polarity patch.  In describing the magnetic setup, for simplicity we will assume 
that the majority-polarity field is largely open into the heliosphere, such as in a coronal hole.  The same arguments apply for 
closed-field regions, as is typical of quiet Sun and active regions (such as those studied in this paper), as long as the far end 
of the field in the region extending beyond the null is far enough away so that the jet can form before it feels substantial physical 
effects of the closed-loop's far end.  Prior to eruption the minifilament (consisting of chromospheric-temperature material) sits 
in the sheared-field core of a magnetic-arcade field that consists of magnetic loops 
connecting the minority-flux patch and the surrounding majority flux, on one side of the minority patch  
\citep{adams.et14,panesar.et16a},
as in the schematic drawn in \citet{sterling.et15} \citep[and also in slightly modified form in][]{sterling.et18}.   As the minifilament-carrying
core field, and
its enveloping field, is erupting, that enveloping field reconnects with 
oppositely directed field at the elevated magnetic null; we call this ``external reconnection,'' because it occurs on the outside edge of
the erupting-minifilament's enveloping field. This external reconnection adds new closed-field loops over the non-erupting portion
of the anemone field, leading to increased brightening of an extended part of the anemone in soft X-rays (SXRs) and at
some EUV wavelengths.  That same external reconnection also results in rearrangement and heating of open coronal field on the
far side of the null.  Heated coronal material, and often some or most of the cool minifilament material too, flows out along the newly 
reconnected open field and forms the spire of the jet, while the newly heated loops of the anemone field form a brightened region
at the foot of the jet spire.   In addition to the external reconnection, the erupting field enveloping the minifilament flux rope 
also undergoes reconnection of its legs under the erupting minifilament; we call this ``internal reconnection'' since it happens within the 
erupting anemone-arcade field.  This internal reconnection results in the aforementioned JBP in the base region.  
Generally this is the brightest part of the base region, outshining the external-reconnection-heated anemone loops.  See the schematic 
drawings and accompanying captions in \citet{sterling.et15} (Fig.\,2 of that paper) and \citet{sterling.et18} (Fig.\,1 of that paper) for
more details on this minifilament-eruption process.  Numerical simulations  \citep{wyper.et17,wyper.et18a} and non-linear 
magnetic-field topological modeling \citep{farid.et22} are able to reproduce many of the features of this 
minifilament-eruption picture for jets.

The above description does not address what builds the minifilament and what triggers it to erupt to make a coronal jet.  For
the triggering-to-erupt aspect, investigations show that in many cases, if not most or all cases, {\it magnetic flux cancelation} 
precedes and triggers the eruption.
This has been shown using on-disk AIA images and \sdo/Helioseismic and Magnetic Imager (HMI) line-of-sight magnetograms in a 
number of studies looking at many quiet Sun and coronal hole jets, including \citet{panesar.et16a}, \citet{panesar.et17}, \citet{panesar.et18a}, 
\citet{mcglasson.et19}, and \citet{muglach21}.  \citet{kumar.et19} also look at the magnetic evolution at the base of several jets, 
and concluded that shearing and/or twisting photospheric flows are more common in triggering jet-producing eruptions than cancelation.  
Regarding formation of the minifilament prior to eruptions, \citet{panesar.et17} presented evidence that flux cancelation also 
builds a flux rope that contains the cool minifilament material and erupts to create the jets, in the cases that they examined.

In some cases, the sheared-core/flux rope field might erupt carrying little or no detectable 
cool-minifilament material; e.g., \citet{mcglasson.et19} reported seeing erupting minifilaments in 90\% of the sixty jets that 
they studied, \citet{kumar.et19} report erupting minifilaments in 67\% of their 27 jets, while some other studies \citep[e.g.,][]{sterling.et15,panesar.et16a,panesar.et18a,sterling.et22} report them in essentially all of their jets.  While some of these
differences are likely due to detection thresholds and visual interpretations, it is clear that minifilament eruptions, or
eruptions of flux-rope fields upon which little or no cool material has gathered, is the source of a substantial portion
of solar coronal jets in coronal holes and quiet Sun.  In many cases, these eruptions are triggered by photospheric magnetic
flux cancelation, and flux shearing and/or twisting is likely important in building up free magnetic energy for fueling the eruptions
(and perhaps triggering the eruptions in some cases).

In active regions however, the situation is not as simple for us to summarize.  Because jets have a similar appearance in both active-region 
and non-active-region solar locations, with a long and narrow spire extending out from a bright base location and lifetimes of the order of
$\sim$10\,min, an initial hypothesis is that jets in both types of locations are driven by the same process.  Indeed, this is clearly 
the case for some active region jets (hereafter, AR jets), where an erupting minifilament is clearly seen as forming the jet.  For example, 
in two of our previous investigations on the topic of AR jets, \citet{sterling.et16b} and \citet{sterling.et17}, we found cases where 
AR jets clearly resulted from minifilaments that were erupting.  In other cases, however, \citet{sterling.et16b} and \citet{sterling.et17} found 
it difficult or not possible to identify an erupting 
minifilament as the origin of the jet.  
Those papers did find that
the jets, nonetheless, originated from magnetic neutral lines, and for most of the cases there was magnetic cancelation occurring along those
neutral lines prior to or during the time of jet formation.  Thus, those studies concluded that the same minifilament-eruption mechanism was operating irrespective of whether
an erupting minifilament was discerned. In addition, while the base of most quiet Sun and coronal hole jets show a JBP, that is both
brighter (in terms of intensity per pixel in the images) and compact compared to the overall size of the base region, in AR
jets it can sometimes be difficult to identify a specific JBP brightening, even in cases where an erupting minifilament is apparent.  See 
\citet{hinode.et19} (section on Coronal jets, by A. C. Sterling, in that paper)  for a summary description 
of the difficulties of AR jets in terms of the minifilament-eruption picture. 


There are a large number of AR jet studies in the literature, and we do not address the observations of all of them here.  
We point out that there is no consistency in these studies on whether there is an erupting minifilament at the start of AR jets.   In several cases, 
studies of AR jets make no explicit mention of erupting minifilaments in the context of their observed jets 
\citep[e.g.,][]{mulay.et16,liu.et16a,sakaue.et17,mulay.et17a,mulay.et17b,mulay.et18,miao.et19,odermatt.et22}; the 
focus of many such papers, however, was often something other than the jets' origin, and so we cannot tell whether erupting minifilaments
were not identified because they were not looked 
for at all, or if they were looked for but not found.  (We point out that, in our experience, to see erupting minifilaments in AIA images 
often requires 
dedicated examination of fields of view of $100'' \times 100''$, or even smaller, centered on the site of the jet's origin, and 
high time cadences of $\ltsim$1\,min, or even 12\,s.  Moreover, it can be essential to examine more than one wavelength, as emphasized
in \citeauthor{sterling.et22}~\citeyear{sterling.et22}.  Many studies not having such focused fields of view and high cadences would
miss them.) In some other studies, AR jets are essentially reported as coming from 
erupting minifilaments, consistent with what we have argued for coronal hole and
quiet Sun jets \citep[e.g.,][]{bhuwan.et16,hong.et17,doyle.et19,solanki.et20,zhang.q.et21}.  
Yet some other works describe AR jets
as having a {\it geyser} appearance \citep{paraschiv.et19,paraschiv.et20,paraschiv.et22}.  These works are interesting in regard to
erupting minifilaments in that they find the jets to be consistent with forming from such minifilament eruptions, although there is 
no overt reference to them being clearly observed at the base of the geyser jets.  Other studies \citep{joshi.r.et17,chandra.et17} 
also observe AR jets that the authors state are consistent with the minifilament eruption idea, but which do not make explicit mention 
of having observed an erupting minifilament at the base of the specific AR jets that they study.  And some studies find (mini)filament eruptions
but argue that those eruptions do not cause the AR jet \citep[e.g.,][]{joshi.r.et20}.  Therefore, overall, these other studies appear to be
consistent with some AR jets originating from obvious minifilament eruptions, while in other cases the existence of minifilament eruptions in the 
production of AR jets is, at best, less obvious.

In this paper, our objective is to address the question of whether AR jets are made by minifilament eruptions, even when the jets appear
as geysers without an obvious visible erupting minifilament.  We have found an AR that has jets in two locations, on the region's south side and
on its northwest side.  We will show that the south-side jets are formed by the picture we present above: erupting minifilaments that are triggered
to erupt by flux cancelation.  The northwest-side jets, in contrast, appear geyser like without an obvious erupting minifilament visible.  We argue
that, under the assumption that jets in both locations were caused by the same mechanism, the jets would have a different appearance merely 
as a consequence of our viewing perspective: The south-side jets originate from a location where we have an unobstructed view to the origin 
sight, allowing us to see the erupting miniflaments.  The magnetic field of those south-side erupting minifilaments, however, reconnects with 
far-reaching coronal field before erupting very far, and that reconnection simultaneously destroys the minifilament and gives the jet its 
geyser-like appearance.  For the northwest-side jets, however, our view of the jet-origin site is obscured, likely by cool and dense 
chromospheric-like material suspended in the low atmosphere of the AR; we argue that this obscuration would prevent our detection of 
erupting minifilaments even if they were initially present in those northwest-side jets.  With this, we offer a possible explanation for why some
AR jets appear to occur without minifilament eruptions, even if that is what made them.  We are not able to {\textit prove} that 
minifilament eruptions make those jets, but we argue that the scenario is reasonable and plausible.

A note on terminology:  \citet{sterling.et15} used the term ``minifilament" to describe the features that erupted to make the jets they 
observed, because those features had a filament-like appearance in EUV images, but were about an order-of-magnitude smaller in extent
than the (3---11)$ \times 10^4$\,km  range given for filament lengths by \citet{bernasconi.et05}.  AR filaments, however, might be smaller than 
those in the \citet{bernasconi.et05} survey.   We use the term ``minifilament" here to mean basically the ``small-scale filament-like 
features that erupt to make jets," even in the AR setting.

\comment{
Here we examine a set of AR jets that provides fresh insight into the question of whether minifilament eruptions drive
them.  In this case, the same active region expels several AR jets from two different locations in the active region.  One 
of these locations is to the south of the region's sunspot, and the second location is on the northwest side of the sunspot.  The viewing 
perspective is such that we can observe the start of the south-side jets from nearer their actual origin site than we can observe 
the northwest-side jets.  In the south location, we can see clearly an erupting minifilament in the base of the jets.  For the 
northwest-side location, 
there is no clear indication of such erupting minifilaments; we argue, however, that even if an erupting minifilament makes those
jets too, our view of the erupting minifilaments would be obscured by foreground solar-atmospheric material.
We suspect that the same basic process makes jets in coronal holes, quiet Sun, and in active regions.  But the  prevalence of 
such foreground material in active regions, and the apparent low altitude at which the ``filament character" of the erupting minifilament disappears,
likely exacerbate the difficulties in detecting erupting minifilaments in the source location of many AR jets.

} 

\section{Instrumentation and Data}
\label{sec-data}

We use UV and EUV images from the Atmospheric Imaging Assembly \citep[AIA;][]{lemen.et12} 
on the Solar Dynamics Observatory (\sdo) satellite.  AIA nominally observes the full Sun in 
seven EUV filters, of characteristic wavelengths of 304,  171, 193, 211, 131, 335, and 94\,\AA, each 
at 12\,s cadence.  And it also nominally observes in two UV channels, at 1600 and 1700\,\AA,
at 24\,s cadence.  All of these are images are obtained with detectors with square pixels of width 
$0\arcsec\kern-0.5em.\,6$.  We also use mainly line-of-sight magnetograms from \sdo's 
Helioseismic and Magnetic Imager \citep[HMI;][]{scherrer.et12}. These are obtained with a cadence
of 45\,s, and with square pixels of width $0\arcsec\kern-0.5em.\,5$.  We cross check the validity of
these line-of-sight magnetograms with an HMI vector magnetogram.

We observe the active region NOAA AR~12824, on 2021 May~22.  This was an extremely jet-rich region
over an extended period, showing activity May 21---24.  Here we concentrate on a subset of this period,
covering 2021 May~22, over 0---8\,UT\@.  For this eight-hour period we have inspected all nine of the 
EUV and UV AIA channels at full cadence.  This region was also responsible for expulsion of a series
of $^3$He-rich solar energetic particle (SEP) events, with the jets discussed here likely being
the sources of those SEPs \citep{nitta.et23}.

Figure~\ref{hide_zu1} shows an overview of the region in selected AIA channels, during the time of a south-side jet.  
All six panels are at about the same time. Panel~(a) shows an 
AIA\,1600\,\AA, and panel~(d) shows the same 1600\,\AA\ image with HMI magnetic flux contours 
overlaid.  This AIA channel is of \civ\ lines and nearby continuum, and shows photospheric features. 
Panel~(b) is a 304\,\AA\ image, formed primarily from \heii\ and showing chromosphere and transition region
emissions.  Panel~(c) is a 171\,\AA\ image, from \feix\ and showing cooler corona and hotter transition region
emissions.  Panel~(e) is a 193\,\AA\ image, which shows \fexii\ emission near 1.6\,MK, but with 
additional contribution from \fexxiv\ near 20\,MK during strong flaring times.  Panel~(f) is a 94\,\AA\ image,
which is from \fexviii\ at around 6.3\,MK, although this 94\,\AA\ channel also contains cooler emissions
too \citep[][]{lemen.et12,warren.et12}.  Except for the 193\,\AA\ channel at the brightest flaring locations, the 94\,\AA\ 
channel generally shows the hottest general coronal emissions in our presentations in this paper.  Panel~(a) 
shows a dominant sunspot in the observed field of view.  At the time of these images a strong event is occurring 
that is visible in all panels on the south side of that spot. The EUV channels show that material is ejected out from the west side 
of that brightening, roughly collimated and directed toward the south; this is the spire of a strong 
AR jet \citep[e.g.,][]{sterling.et17}.

Figure~\ref{hide_zu2} has the same arrangement as Figure~\ref{hide_zu1}, but this time for when an event is occurring in the second 
very active location of this region, to the northwest of the sunspot.  Again the EUV images show a jet spire being expelled
from the brightening, this time directed toward a north-westward direction.

An animation accompanying Figures~\ref{hide_zu1} and~\ref{hide_zu2} shows the evolution of the region over our eight-hour 
time span.  There are repeated eruption and jetting episodes over this period, with the major such episodes 
occurring at or near the two locations highlighted in those two figures, i.e.\ in the south and in the northwest.

Figure~\ref{hide_zu3} shows lightcurves from the two regions.  Panel~(a) shows the \goes\ soft X-ray intensity changes over
the discussed time  period.  Panels~(b) and~(c) are from the AIA~94\,\AA\ channel, and were obtained by integrating
that channel's intensity over the region of the turquoise box in Figure~\ref{hide_zu1}(c) for the southern-location lightcurve in 
Figure~\ref{hide_zu3}(c), 
and over the green-box location in Figure~\ref{hide_zu2}(c) for the northwestern location lightcurve in Figure~\ref{hide_zu3}(b). 
{By using the 94\,\AA-channel lightcurve, we are selecting out only the relatively hot (and hence, energetic) events for the regions; from
the 94\,\AA\ video accompanying 
Figures~\ref{hide_zu1} and~\ref{hide_zu2}, it is apparent that the strong brightenings tend to occur at localized locations, and so the
lightcurve of Figure~\ref{hide_zu3}(b) is from primarily those localized locations; that is, the intensity peaks in
the plot are from those brightenings.  (This is similar to the situation with the \goes\ profile in Fig.\,\ref{hide_zu3}(a); \goes\ observes
the entire solar disk, but the strong X-ray emission peaks at any given time
mainly come from localized bright flaring regions.) That same movie shows that at the time of the two intense brightenings in the 
southern location near 3\,UT and
near 6:20\,UT, bright diffraction spikes spread scattered light into the Figure~\ref{hide_zu2} green-box location.  This results 
in faux intensity peaks in the intensities of these times plotted in Figure~\ref{hide_zu3}(a), and we indicate that
these are not real with shaded blue rectangles in that plot.  These two events from the south region each produced flares
of the \goes~C6 level.  There are also two relatively large flares in the northwest region, with the first peaking near 3:35\,UT and reaching 
a \goes\ B3 level, and the second of these peaking near 
6:50\,UT at the \goes\ C1 level.  We will call the eruptive events making these four flares ``main eruptions," with two main eruptions
occurring in each of the locations.   From
the 94\,\AA\ lightcurves, there are also three smaller flares before the south location's first main eruption, and one smaller flare before
the second main eruption, with all four of those smaller flares being at the \goes\ B1---B2-level.  We will call these four smaller flares ``precursor
events" to the main eruptions, for reasons discussed below (at the start of \S\ref{sec-results}).  Table~\ref{tab:table1} summarizes the primary
events detected over our eight-hour observation period.

Figure~\ref{hide_zu4} shows an HMI magnetogram of the active region.  The sunspot, of positive (white) polarity, dominates
the region.  It is surrounded, however, by copious negative (black) flux.  The animation accompanying the figure shows that
the much of this negative flux, and some positive flux as well, flow outward from the spot with time, as is typical with 
moving magnetic features (MMFs).  In contrast to those outflows, a near-vertically oriented ridge of positive flux,
between roughly $-430$ and $-440$ on the abscissa, is nearly stationary relative to the spot over the observation period.
Contours in Figure~\ref{hide_zu4}(b) are the same as those used in Figures~\ref{hide_zu1}(d) and Figure~\ref{hide_zu2}(d).
Figure~\ref{hide_zu4}(c) focuses in on the southern location, which is denoted by the blue box in (b).  The displayed magnetogram 
in Panel~(c) saturates at a lower field strength than those in the first two panels, to show more readily some of the 
comparatively weaker fluxes that appear to be critical to some of the eruptions, as we will discuss further below.

Because the magnetogams displayed in Figure~\ref{hide_zu4} are line-of-sight obtained from when the location is moderately
away from disk center, $\sim$(20$^\circ$\,N, 15$^\circ$\,E), there could be a concern that projection
effects result in differences between the line-of-sight field compared to the actual radial magnetic field.
To check this, we have also created a magnetic field map of the same region and time as in Figure~\ref{hide_zu4}, using a disambiguated vector radial-field component HMI vector magnetogram.  Specifically, we use the HMI ``SHARP" 720\,s 7590 patch, CEA projection (where the magnetogram is 
remapped to a heliographic Cylindrical Equal-Area, CEA, projection centered on the patch; \citeauthor{bobra.et14}~\citeyear{bobra.et14}).
We show this in Figure~\ref{hide_zu4a}.  This shows that the fluxes in the south region are in qualitative agreement with the line-of-sight 
versions over roughly the area of the blue box in Figure~\ref{hide_zu4}(b), and with the comparatively weak negative flux in the yellow box in  Figure~\ref{hide_zu4} visible in
the orange box in Figure~\ref{hide_zu4a}.  For the northwest fluxes however, the deprojected vector magnetogram in 
Figure~\ref{hide_zu4a} shows more negative-polarity flux than does the line-of-sight magnetogram over roughly the green box of Figure~\ref{hide_zu4}(b).  Therefore,
we exercise caution not to over-interpret the northwest fluxes in building our case.  We still, however, will rely on the line-of-sight 
magnetograms for our main analysis, because the vector magnetogram’s detection of transverse field strength $\ltsim$100\,G is
not reliable (\citeauthor{paraschiv.et20}~\citeyear{paraschiv.et20} found a mean noise threshold value of $\sim$130\,G for the transverse
field for their near-limb region), and the deprojected radial field of Figure~\ref{hide_zu4a} is a composite of the line-of-sight and transverse fields.  (We will discuss 
other vector field observations of jet-producing regions in \S\ref{sec-discussion}.)

In addition to our two selected jetting locations that are on the south side and on the northwest side of the AR, there are at least two additional 
jetting locations in the region.  Both of these are active simultaneously over 03:49---03:53\,UT,  at location (-390,320) for one of them 
and at location (-380,340) for the other one,  in the video accompanying Figures\,\ref{hide_zu1} and \ref{hide_zu2}
(and they are even better seen in Fig.\,\ref{hide_zu5} below).  But we focus only on the jets that we have identified in Table~\ref{tab:table1} because: 
(a) the main eruptions among the Table-\ref{tab:table1} events are more energetic than those of the other two regions; (b) they are larger in physical size than those in the other two
regions, allowing us a better chance to resolve the onset process(es); and perhaps most importantly (c) the jets we identified in what we call
respectively the southern region and the northwest region are two contrasting 
cases, with the jet-origin site visible to us in the southern one and the jet-origin site obscured from us in the northwest region.  Focusing only
on these two locations allows us most simply and straightforwardly to make our main point of the paper: that the minifilament-eruption process
making the jets in the southern region likely also makes the eruptions in the northwest region, but the northwest erupting minifilaments  plausibly 
are not visible due to obscuration of that location along our line of sight.

\section{Results}
\label{sec-results}

We present results of more detailed investigations of the jet-like events in the two regions of AR~12824, over our
observations period of 2022 May~22 0---8\,UT\@.   In Table~\ref{tab:table1}, events~4 and~7 are south-location main events, and 
events~1, 2, and 3 are precursors to the first main event and event~6 is a precursor to the second main one.  In the northwest 
there are two main events: events~5 and~8.  

We group events that originate from
the same location on the same neutral line into ``precursor"
and ``main."  That is, the south-location events~1---4 all originate (as identified by the earliest brightenings in the AIA images) from 
the same magnetic neutral line 
location.  Events 6---7 are also in the south location, but their origin location is north of the origin location of events~1---4; we show the
origin location of events~1---4 in the yellow box in Figure\,\ref{hide_zu4}(c), and we will show the origin location of events~6 and~7 in
Figure\,\ref{hide_zu9}(a) (below).  It is clear that the northwest events 5 and~8 are on a different neutral line from the south-location 
regions.  Because the final jets in the south-location groups, i.e.\ events~4 and~7, are larger than the earlier events from the same 
neutral-line location, we call them main events, and the earlier ones precursor events.  We expect that all of these events operate 
in essentially the same fashion, with the main ones being more explosive then the precursor ones, similar to how standard and blowout 
jets operate in the same fundamental fashion but with the blowout ones generally more explosive than the standard ones 
\citep[e.g.,][]{sterling.et22}.  We concentrate our discussion on these 
eight events.  We consider the two locations separately, first the southern location and then the northwest location.


\subsection{Southern Region: First main event}
\label{subsec-south}

We first focus on the jets originating from the southern location, indicated by the blue box in Figure~\ref{hide_zu1}.
Figure~\ref{hide_zu5} shows frames from that box, with the same layout as Figures~\ref{hide_zu1} and ~\ref{hide_zu2}.
The green arrow shows that there is an absorption feature, which we will shortly argue (Fig.~\ref{hide_zu6}) is an erupting 
minifilament, appearing near a neutral line indicated by the yellow arrows in \ref{hide_zu5}(d).  The magenta arrows 
point to a feature that appears as the west side of a circular ribbon in the 1600\,\AA\ frame (\ref{hide_zu5}(d)), and 
the there is a corresponding circular mound of brightening in the other channels in the figure, with the magenta 
arrows of \ref{hide_zu5}(f) pointing to the west side of that bright mound in an 94\,\AA\ image.  From \ref{hide_zu5}(d), this 
circular-ribbon/bright-mound feature is located in the positive polarity surrounding 
an island of negative polarity.  Since the sunspot is positive polarity, the negative island forms a jet-like setup, with 
a minority-polarity island surrounded by majority polarity, forming an anemone region.  The circular ribbon and mound
of illumination of Figures~\ref{hide_zu5}(d) and~\ref{hide_zu5}(f) are rooted in the positive polarity surrounding 
that negative island.  (Such circular or semi-circular
patters from reconnection in jets is discussed in \citeauthor{sterling.et16b}~\citeyear{sterling.et16b}.)

\comment{
Moreover, the strongest ribbon brightenings are at the locations where the positive flux clumps 
are comparatively strong.
} 

Figure~\ref{hide_zu6} shows 304\,\AA\ frames from the same FOV as in Figure~\ref{hide_zu5}.  From \ref{hide_zu6}(a) 
and \ref{hide_zu6}(b), the minifilament (green arrows in 02:42:41\,UT frame in \ref{hide_zu6}(b)) apparently erupts 
from the edge of the positive field pointed 
to by the yellow arrows in Figure~\ref{hide_zu6}(a).  We first see this feature distinctly at about its location in 
Figure~\ref{hide_zu6}(b), when it appears ``filament-like,'' i.e.\ with both ends rooted in the photosphere, a few arcsec 
southeast of the
lower green arrow in (b).  By the time of Figure~\ref{hide_zu6}(d) however, this minifilament no longer appears tied
to the photosphere at its southeastern end.  Instead, by this time it has become part of the jet spire, whipping around
toward the west in the animation accompanying Figure~\ref{hide_zu5}.  The distance traversed between when it looks
like a minifilament and when it ``loses its minifilament character'' is about the east-west distance between the tip of the
lower green arrow in \ref{hide_zu6}(b) and the tip of the blue arrow in  \ref{hide_zu6}(d).  The tips of these arrows 
are projected onto the abscissa in \ref{hide_zu6}(b), and the distance between them is about $3''$; if we assume that
the erupting minifilament travels an equal distance in the south direction also, then the total distance displaced 
is about $4''$.

We can examine in more detail the magnetic character of the location from which the minifilament erupted, using 
Figures~\ref{hide_zu6}(a) and~\ref{hide_zu4}(c).  Figure~\ref{hide_zu6}(a) is at the same time as Figure~\ref{hide_zu6}(b)
and thus shows the location on the magnetogram near the time when the erupting minifilament first becomes 
clearly visible.  From the magnetogram
contour in Figure~\ref{hide_zu6}(a) this location seems to be just a ridge of positive flux.  But in the deeper magnetogram
of Figure~\ref{hide_zu4}(c), along with the accompanying animation, there is opposite polarity negative flux present too, albeit
comparatively weak relative to the nearby positive flux.  From the animation, the negative flux flows toward the positive flux
over the first few hours of our observation period, and apparently cancels at the neutral line between them.  

We show evidence for this cancelation in Figure~\ref{hide_zu7}, which plots the total negative flux integrated over the yellow box in Figure~\ref{hide_zu4}(c).  (Specifically, we sum over the negative flux of strength $\gtsim$10G contained in the yellow box in Fig.\,\ref{hide_zu4}(c), 
at each timestep.) We selected this box location to be approximately centered on where event-4's pre-eruption minifilmament 
first becomes visible over about 2:41---2:42\,UT in the 304\,\AA\ images, and we want to observe whether there are magnetic
changes there during the time of the minifilment's formation and eruption start.  We only track the 
negative flux, because, as can be seen from the animation accompanying Figure~\ref{hide_zu4}(c), the negative flux in that yellow-boxed
location at the start of the period, and/or the negative flux that develops (via emergence or via coalescence of beyond-detection 
negative-flux elements) within that box over the period, is largely confined to within that box,
except for some black flux that enters into the north part of the box from about 5:40\,UT\@.  In contrast, the positive flux spills
across the box's boundary on the north, south, and east, and therefore we are not able to track the positive flux changes within the box with 
confidence.  We start tracking from 22\,UT on 21 February so that we can capture the peak relevant negative flux that 
near 23\,UT on that date. We ignore fluxes of absolute value less than 10\,G, to avoid background noise in the flux.  Figure~\ref{hide_zu7} shows that this negative flux peaks at about $5\times10^{18}$\,Mx, and shows
a definite and essentially monotonic decline from the start of the period until just prior to 3\,UT\@.  Four vertical lines show the times
of events~1---4 of Table~\ref{tab:table1}, with the first three precursors in green and the main eruption in red.  This flux decrease 
continues over the period covering all four of these eruptions.  In total, about $1\times 10^{19}$\,Mx cancels over this period, 
assuming equal amounts canceled for both negative and positive flux.   There also is a flux increase starting from around 5\,UT 
that corresponds to the flux that entered the northwest portion of our computation box, but its amount is low compared to the flux 
change over the first three hours of the period.  

\comment{

Another factor in this main eruption, however, is a nearby eruption that starts a few minutes prior to the main eruption, from about 2:38\,UT\@.
This is at the location shown by the black arrows in Figure~\ref{hide_zu6}(b).  This eruption also results from an erupting filament from
a neutral line.  Therefore, it is plausible that the flux cancelation at the location indicated by the yellow box in Figure~\ref{hide_zu4}(c)
prepared the minifilament there for eruption, and then the eruption at the location indicated by the black arrows in Figure~\ref{hide_zu6}(b)
finally triggers that main eruption (in the yellow box) to occur.  The circular ribbon and circular mound of emission (the west 
side of which is pointed to by
the magenta arrows of Figs.~\ref{hide_zu5}(d) and~\ref{hide_zu5}(f)) would be expected to from external reconnection at the null
at the top of the anemone in the minifilament-eruption picture described in \S\ref{sec-introduction}.

} 

These observations suggest that the scenario outlined in \S\ref{sec-introduction} for jet production via minifilament eruption, with 
flux cancelation building the minifilament field that holds the cool minifilament material, is operating here also, to form the minifilaments
that erupt in the making of jets in the south location over the first three hours of our observation period.  This flux amount of 
$\sim 1\times 10^{19}$\,Mx is somewhat higher than that which we have observed in non-AR jets.  For example, 
\citet{panesar.et18a} found the canceled flux below minifilaments that erupted to make coronal hole jets to be 
$\sim$0.5---2.0$\times 10^{18}$\,Mx, and \citet{panesar.et16a} found the canceled flux below minifilaments that erupted to 
make quiet Sun jets to be 
$\sim$0.9---4.0$\times 10^{18}$\,Mx.  Our value here for the flux canceled below a minifilament that erupts to make AR jets, 
while being above the flux that cancels for for coronal hole and quiet Sun jets, is lower than the flux that we found to cancel 
along neutral lines and make CME-producing
eruptions in small active regions, which is around $10^{20}$\,Mx (see Table~1 in \citeauthor{sterling.et18}~\citeyear{sterling.et18}).   
From the location in the yellow box of the animation accompanying Figure~\ref{hide_zu4}(c), the cancelation of the
negative-polarity flux patches occurs over $\sim$1---3\,UT, which corresponds closely to the time of jet events 1---4 (Fig.\,\ref{hide_zu2},
bottom panel), which originate from that neutral line. Moreover, after the final traces of those negative-polarity patches have canceled,
the jetting from that location ends.  This supports that the cancelation is the cause of the jets from that location.  This is similar to the
case for active region jets in \citet{sterling.et17}, where jetting at specific locations ended after the cancelation episodes at that location 
ended; it is also similar to the situation with homologous quiet Sun jets discussed in \citet{panesar.et17}, where the homologous jets 
continued until the minority-polarity flux at the base of the jetting neutral line completely disappeared.
Therefore our results here are consistent with other results, in indicating that flux cancelation occurred at the minifilament-eruption 
location in the hours prior to the AR jets, and triggered the eruptions that made those jets (referring to events 1---4 in Table~\ref{tab:table1}).  
After the flux cancelation 
ended, there were no further jetting events from that location.  The eruption at the location of the black arrows 
in Figure~\ref{hide_zu6}(b) occurred slightly before event~4, and likely abetted that main eruption.

\subsection{Southern Region: First-Main-Event Precursors}
\label{subsec-south_events}

Next we consider the three precursor events prior to the first main event; these are events 1---3 in Table~\ref{tab:table1}, and the times
of which are indicated by the green lines in Figure~\ref{hide_zu7}.  From the AIA~94\,\AA\ animation accompanying Figures~\ref{hide_zu1}
and~\ref{hide_zu2}, each of events~1---3 were located
at the site of the main eruption, and the animations of the other EUV channels suggest that all three events form in basically the same
fashion.

Figure~\ref{hide_zu8} shows the third of these precursor events.  It shows that the source of this precursor eruption is the 
same minifilament field (more precisely: the field in and enveloping the minifilament that eventually erupts to 
make the main jet event~4) as that which 
launched the first main eruption, event~4.  Here we say that it is the ``same minifilament field" instead of the ``same minifilament," because
it is unclear whether part of the exact same filament-like feature is erupting outward in both cases, but it is virtually certain that strands
of the same magnetic field structure are erupting, since both occur in the same vicinity along the same neutral line.  Moreover, 
Figure~\ref{hide_zu7}, along with the animation accompanying Figure~\ref{hide_zu4}, shows that flux cancelation is continuing 
along that same neutral line throughout the occurrence of all three of the precursors and the first main event.

 From Figure~\ref{hide_zu8} and the animation accompanying Figure~\ref{hide_zu4}, strands of the minifilament field at the location 
 indicated by the yellow arrows in Figure~\ref{hide_zu8} are pealing off and erupting, with the green arrows in Figure~\ref{hide_zu8} pointing 
 to one of these strands.  In this case the strand has a width of only $\sim$1---2$''$, and the first two precursors show erupting 
 strands of similar widths.  In contrast, the first-main event (event~4) had a minifilament whose width varied in the EUV images, but 
 was $\gtsim$5$''$-wide at times (such as in the 2:46:53\,UT 304\,\AA\ frame in the video accompanying Figure~\ref{hide_zu4}).  In the second 
 and third precursors material eventually flows out along jet spires, similar to the case with 
 the erupting minifilament that made event~4, and after they traveled about the same distance from the
 site from which they were first detected as in the event-4 case (i.e., $\sim$5$''$).  It is not clear whether the minifilament strand of the first precursor undergoes a
 complete eruption along the jet-spire field, in which case it may be of the confined-eruption variety \citep[e.g.,][]{sterling.et22}.
 
 Because the flux cancelation is continuing during the period of the precursors, our observations suggest that the continued flux cancelation
 builds up the minifilament, presumably adding shear to the minifilament field. This leads to repeated partial releases of the built-up
 shear along the minifilament through the precursor eruptions (events~1---3).  Finally, enough cancelation occurs and builds up enough 
 shear in the
 minifilament-flux-rope field that the field becomes unstable and ejects outward, producing the first main eruption (event~4).  
 
 \comment{
 
 This final eruption 
 may have been abetted by the eruption indicated 
 by the black arrows Figure~\ref{hide_zu6}(b) along a nearby neutral line.

} 

\subsection{Southern Region: Second-Main Event}
\label{subsec-south_second}

We now look at the second main event in the south location, event~7 in Table~\ref{tab:table1}.  
Both this event, and its precursor (event~6; \S\ref{subsec-south_second_precursor}), are included in the animation 
accompanying Figure~\ref{hide_zu5}.

Figure~\ref{hide_zu9} has a four-panel layout similar to that of Figures~\ref{hide_zu6} and~\ref{hide_zu8}, but for event~7.  In
panels~(a) and~(b) the yellow arrows point to a minifilament that eventually erupts to create the jet of the event.  As with 
the events~1---4 cases, this minifilament is not on a neutral line that is well defined in the contour levels in panel~(a), but comparing
with Figure~\ref{hide_zu4}(c) and its accompanying video shows that there are many relatively weak-level mixed-polarity elements in the
region, and these show dynamic movement throughout the video.  This minifilament forms at a location north of the location of 
events~1---4 that has substantial intermixing of fluxes, with many flux cancelations likely occurring.  

In Figure~\ref{hide_zu9}
we point out two locations that are good candidates for the source location. One is the set of isolated positive-polarity patches embedded
in a sea of negative polarity (black arrow in Fig.\,\ref{hide_zu9}(a)), that shrinks and partially disappears near the time
of the jets, as can be seen by cross-identifying the same polarity patches in the animation accompanying Figure~\,\ref{hide_zu4}(c).  The second location is
the neutral line just west of the yellow arrow in  Figure\,\ref{hide_zu9}(a) (below); from the animation accompanying Figure\,\ref{hide_zu4}(c),
negative flux appears to flow eastward and positive flux appears to flow westward with time, converging on that neutral line.  These locations
are extremely close to the location of the very earliest brightenings in the onset of events~6 and~7, with the times of the panels 
in Figure\,\ref{hide_zu9} very near the time of the earliest brighting of event~7.   Both of these locations are
candidates for flux cancelation that could result in minifilament eruptions causing the second event and its precursor, although we
are not able to confirm definitively that these cancelations are the source of those events.  
These canceling polarities are not as well isolated over the entire period of the second set of eruptions as are the key canceling polarities
of the first set of eruptions, and therefore it is not possible to show the flux changes in the manner of Figure\,\ref{hide_zu7} for 
this second-erupiton case.

The filament-like feature
has started moving toward eruption in panel~\ref{hide_zu9}(c), leaving a brightening adjacent to it at its original location, and this
would correspond to its JBP.\@  At this stage the feature still has its ``filament" character, but by panel~\ref{hide_zu9}(d) it has started
to flow outward along far-reaching field lines, along the jet's spire.  Similar to our procedure for event~4, we can measure the distance that the
minifilament moves from where we first detect it near its location in panel~\ref{hide_zu9}(b), to when it no longer has a filament
character in panel~\ref{hide_zu9}(d).  That distance is about $4''$ in the east-west direction, or $\sim$6$''$ if an equal southward 
displacement is assumed.

\subsection{Southern Region: Second-Main-Event Precursor}
\label{subsec-south_second_precursor}

A precursor, event~6 in Table~\ref{tab:table1}, occurs prior to the second main event (event~7) in the south location.
From the table, and the animation accompanying Figure~\ref{hide_zu5}, this precursor event peaks in 94\,\AA\ 
intensity at 5:40\,UT\@.  That same animation shows that the locus of the event's origin is the same as that of the
following second-main eruption, event~7.  Moreover, just as precursor  events~1---3 to event~4 show minifilament strands
erupting out of the location where the larger minifilament erupts in event~4, in this case a distinct minifilament strand is
apparent erupting out of the location where a larger minifilament erupts to make event~7.  For this case the minifilament 
strand is visible in the 304\,\AA\ movie from about 5:24\,UT, and by about 5:35\,UT it has started to move out along the same
far-reaching field on which the second-main-event's (event~7) spire forms.  Accompanying its eruption is a 
JBP brightening from where the minifilament resided prior to eruption, visible over about 5:25---5:54\,UT
in 94\,\AA, and at similar times in other channels of the Figure~\ref{hide_zu4} animation.  The distance traversed by the 
erupting-minifilament strand between the time of its filament-like character (rooted at both ends to the photosphere) and
the time it loses that minifilament character as it flows out along the jet-spire field is about the same as for the 
case of \S\ref{subsec-south_second}, i.e.\ $\sim$5$''$.  As with the precursor events 1---3 to event~4, the 
erupting-minifilament strand for this precursor (event~6) is also of width $\sim$1---2$''$.

Therefore, this precursor (event~6) to the south location's second main eruption (event~7), is analogous to 
the three precursors (events~1---3) to the south location's first main eruption (event~4).

\subsection{Northwest Region}
\label{subsec-northwest}

Next we consider activity in the northwest region.  Figure\,\ref{hide_zu10} shows the region with the FOV of the green box
in Figure~\ref{hide_zu2}(c), at a time of event~5 of Table\,\ref{tab:table1}, the first of two main events that we list for this region.
The green arrow in Panel~\ref{hide_zu10}(b) shows absorbing material in the process of being ejected from the region 
along a jet spire.  From the
video accompanying the figure, this material originates from below more-slowly-moving absorbing material, with the two
yellow arrows in \ref{hide_zu10}(b) pointing to two examples of this absorbing material.  There is much of this absorbing material present, 
and it acts to conceal 
the locations from which the material ejecting along the jet spire (green arrow) originates.  Based on our previous jet studies, and
based on the events we examined above from the south location for this region, it is plausible that this ejecting absorbing 
material originated as a minifilament that was expelled from a magnetically mixed-polarity location.  
(Fig.\,\ref{hide_zu4a} confirms that the mixed polarity in this location persists even in the disambiguated radial-component 
vector map.)

Orange arrows in Figure~\ref{hide_zu10}(d) point to candidate mixed-polarity locations from which the absorbing material 
may have existed as a minifilament prior to its eruption.  These locations show enhanced UV brightenings in this figure, and also
in several different EUV wavelengths in this figure and in the video accompanying Figure~\ref{hide_zu10}.  Thus, one or more of 
these locations are consistent with being a JBP made under an erupting minifilament.  Each of these orange-arrowed locations is $\sim$30---40~arcseconds away
from the exposed portion of the material flowing out along the jet spire pointed to by the green arrow in Figure~\ref{hide_zu10}(b).  From our
examinations of the erupting minifilaments in the south location, we saw clear evidence for erupting minifilaments to 
make the material that flowed out along the respective jet spires, but those erupting minifilaments lost their filament 
character after moving only $\sim$5$''$ from the mixed-polarity locations from which they were expelled.  So if the same
thing happened in the northwest-event case of Figure~\ref{hide_zu10}, then the pre-erupting and erupting minifilament
would not be visible over such as short distance from the expulsion locations.  Thus, the cool material of the erupting minifilament
would first appear as an outward-moving clump of jet-spire material, exactly as we observe in Figure~\ref{hide_zu10}.

There are several similarities between the events of this northwest location and those in the south discussed earlier.  
As discussed in \S\ref{subsec-south}, during the events in the south region a distinct circular ribbon appeared as 
the minifilaments were erupting and material was flowing out along the jet spire, with the magenta arrows in Figure~\ref{hide_zu5}(d)  
pointing to the west side of one such ribbon, and the magenta arrows in  Figure~\ref{hide_zu5}(f) pointing to corresponding 
mound of EUV brightenings.  A similar such feature is apparent in the northwest-location events too.  In  Figures~\ref{hide_zu10}(d)
and~\ref{hide_zu10}(f), the magenta arrows point to the west side of a structure that, together with the orange arrows, forms a
partial-circular structure.  As in the Figure~\ref{hide_zu5} case, the ribbons and brightenings are strongest at locations where
the magnetic flux clumps are strongest.  

Another similarity with events in the south location is that again there is an region with
included negative-polarity negative field that is (partially) surrounded by dominant-polarity positive field.  In 
Figure~\ref{hide_zu10}(d), there is positive field (red contours) partially surrounding negative (green contours)
field, and the partial-circular ribbon is largely rooted in that positive field.  Thus, the magnetic setup here is that
of an anemone region, just as it is in the south location also.  The partial-circular ribbons and brightenings of
Figure~\ref{hide_zu10} are thus also likely due to the eruption of a minifilament, where the field enveloping the cool
minifilament material undergoes external reconnection with far-reaching field at the magnetic null at the top of the 
anemone.  This results in newly reconnected closed-loop field rooted in the locations indicated by orange and
magenta arrows of Figure~\ref{hide_zu10}(d), while the cool filament material flows out along the jet spire 
on far-reaching field, as is happening with the material pointed to by the green arrow in Figure~\ref{hide_zu10}(b).
Although the line-of-sight magnetogram used here likely underestimates the negative flux in the northwest
region (Fig.\,\ref{hide_zu4a}), the vector-magnetogram-derived radial field still shows substantial mixed polarity
in the northwest location, which plausibly cancels to make the jet.  We can say that the locations indicated by the 
arrows in Figure~\ref{hide_zu10}(d) are candidate locations for flux cancelation leading to minifilaments that 
erupt to cause the observe northwest-side jets.

Event~8 of Table~\ref{tab:table1} is the second main event from the northwest location.  From the animation 
accompanying Figure~\ref{hide_zu10}, this event is morphologically similar to the just-discussed event~5 in the
same location.  Thus these two events, events~5 and~8, are roughly homologous, in a manner similar to events~4 and~7 in the
south location.

The flare accompanying the eruption producing this jet (event~8) is of \goes\ C1 level, and thus more energetic than event~5, which
was B4 (see Table~\ref{tab:table1}).  And in this case the partially-circular flare ribbons pointed to in Figure~\ref{hide_zu10}(d) 
for event~5 are even 
more prominent in this case, as can be seen in the video accompanying Figure~\ref{hide_zu10} at about 6:51\,UT\@.

Magnetically, the northwest region around the orange-arrowed sites in Figure~\ref{hide_zu10}(d) are of mixed polarity.  Moreover,
the animation accompanying Figure~\ref{hide_zu4} shows that there is substantial movement and interactions of the mixed polarities
in this northwest location.  Although we cannot pick out a specific location where flux cancelation might be occurring, in a manner
analogous to that which we found in the south location (e.g., Fig.~\ref{hide_zu7}), it is plausible that such cancelation did occur
in that location, creating a minifilament flux rope and contributing to the triggering of its eruption into the jets for events~5 and~8.  The anemone
magnetic setup and the partially-circular ribbons during those events provide additional support for this general scenario.
Any such possible erupting minifilament for these northwest events is largely obscured by overlying and surrounding absorbing 
material, at least before it is being expelled along far-reaching field lines and has lost its minifilament character.  This obscuration 
makes any such erupting minifilament harder to certify than those in the south location, where the obscuration along our line-of-sight
is much less, allowing a less obscured view of the early stages of the jets' onsets.

We recognize that we have introduced speculation about the cause of these northwest-location jets.  This is unavoidable, however, because
the base of those jets is obscured from our view.  We have argued that, based on the distances traveled by erupting minifilaments in making
jets where we can see the jet's base in the south location, a similar minifilament eruption in the northwest would not be visible to us as an
erupting minifilament due to the obscuration.  While we cannot {\textit prove} this scenario for the northwest-location jets, we hold that our
speculation is plausible, given that we can see this process occurring for the south-location jets, and given that we have seen the
same basic minifilament-eruption scenario operating for many jets in coronal hole and quiet Sun locations (see \S\ref{sec-introduction}).

\comment{

Even though we have suggested the plausibility that magnetic flux cancelation was partially responsible for the 
minifilament eruptions to make the jets in the northwest location for events~5 and~8, this cancelation might not be the immediate 
trigger of the eruptions in these cases.  Instead, the immediate trigger could be destabilization of the field holding the 
northwest-location pre-eruption
minifilaments by eruptions that cause the jets in the south location, perhaps after the flux cancelation has partially or largely build
the pre-eruption minifilament.  This is suggested by Figure~\ref{hide_zu3}(b), which 
shows that both events in the northwest region occur shortly following a large jet in the south region, as indicated by the false-intensity
peaks denoted by the blue-shaded boxes in the plot.  These are the result of scattered light from, respectively, events~4 and~7 
entering into the FOV of the 94\,\AA\ intensity box of the north location used to make the Figure~\ref{hide_zu3}(b) plot.   Therefore,
it is likely that those south-location eruptions assisted in triggering the eruption of the minifilaments that made jets~5 and~8 in the 
northwest location of the AR\@.

} 

\section{Discussion}
\label{sec-discussion}

We have examined the cause of some of the extensive jetting activity from  AR~12824, over an eight-hour period on 22 May 2021.  We
can summarize our main findings as follows:

\begin{itemize}

\item All of our jet events of Table~\ref{tab:table1}  originate from mixed-polarity locations.  Each of the two south main events occurs from 
the eruption of absorbing material (e.g., in 304\,\AA) that has the clear appearance of a minifilament, with photospheric roots separated by
$\sim$15---30$''$; this range is large because first it appears as a $\sim$15$''$ filament-like structure (e.g.\ at about 2:42\,UT in 
the 304\,\AA\ animation accompanying Fig.~\ref{hide_zu5}), but it quickly merges with a
second filament-like structure so that the total length is closer to $\sim$30$''$ (e.g., at 2:44\,UT in the just-cited animation).  In each case, the minifilament erupts and makes a jet in a manner consistent with the picture described in \citet{sterling.et15} and discussed in 
\S\ref{sec-introduction}.  

\item The three south-region precursors (events~1---3) for the first main event were very similar to that south-region first main event (event~4), 
but the erupting minifilaments were much 
thinner than the main erupting minifilament, and seemed to be merely {\it minifilament strands} from the flux-rope field (or pre-flux-rope 
sheared field) that peel 
off of the larger structure and erupt outward, each making a weak jet and weak flare base brightening (``weak" compared to the
main events).  And then once the larger flux rope (or pre-flux-rope field) erupts in the main event, there are no further jets or 
eruptions from that specific neutral line in the remainder of our observation period.  The precursor (event~6) to the south-location's 
second main event (event~7) was correspondingly similar to events~1---3.

\item In all of these south-location cases (events~1---4, 6, 7), the erupting minifilaments lose their ``filament-like" appearance 
(as absorbing material situated approximately 
horizontally with the solar surface, and appearing to be rooted at two ends in the surface) upon moving only a short distance, 
of $\sim$5$''$ in projection onto the surface, away from the location where it was originally identifiable.  We infer that the 
closed minifilament field reconnects with much-more-extended (far-reaching if not open) field, upon traveling about  5$''$.

\item In the northwest location over our observation period, we identify two events as main northwest events, events~5 and~8.  
In contrast to 
the south-location events, we cannot see a clear indication of an erupting minifilament for either of the two events in the 
northwest region.  From the time of the onset of both events however, we see JBP-like brightenings, indicated by the orange 
arrows in Fig.~\ref{hide_zu10}(d), that could be expected from
a minifilament eruption, are all much greater than $5''$ away from where we first see a clear 
indication of the expelled jet material traveling outward along the jet spire, pointed to by the green arrow in 
Figure~\ref{hide_zu10}(b).  But between the locations of those orange arrows and the green arrow, much of the
area is covered with elevated absorbing material in the EUV panels of Figure~\ref{hide_zu10}.  Therefore, we argue that it is
plausible that the jets of the northwest location did indeed originate from minifilament eruptions, just as did the 
south-location jets (and many, if not most or all, quiet Sun and coronal hole jets; see \S\ref{sec-introduction}),
but the supposed minifilaments in these eruptions are ``hidden" behind that suspended absorbing material, making their detection
difficult or impossible before they lose their minifilament character.

\item For the south-location events we find evidence that 
flux cancelation was an important factor in triggering
the minifilaments' eruptions (Fig.~\ref{hide_zu7}).   For the northwest-location events, we could not unambiguously identify the jet-origin
locations, and therefore we could not confirm that flux cancelation was the primary triggering process for those events.

\item Similar to the south-location jets, the northwest-location jets occur in an anemone region, they show a
circular ribbon (albeit less completely circular than the south-location case), and they likely originate from locations 
with interacting (and likely canceling) mixed-polarity features.  These are all features consistent with
the minifilament-eruption picture for jets \citep[][and discussed in \S\ref{sec-introduction}]{sterling.et15}, 
and thus supports that the jets in the northwest location likely did originate as minifilament eruptions, but the minifilaments
themselves were obscured until their material appeared as outflow on the jet's spire field. 

\end{itemize}

Figure~\ref{hide_zu11} shows a schematic representation of our findings, where a minifilament preparing to erupt and make a jet south of
the spot is unobscured along our line of sight, while a minifilament preparing to erupt and make a jet in the northwest is largely obscured 
along the line of sight by opaque material low in the solar atmosphere of the AR\@. 

Another finding is that, although we could identify a JBP or a candidate brightening
for a JBP in all of our events, these JBP brightenings can often be overwhelmed in brightness by ancillary eruptions
in the region.  AR jets are likely particularly susceptible to such ancillary brightenings, due to the abundance of small-size-scale
but magnetically strong neutral lines in the same AR, which can be triggered to erupt nearly concurrent with the eruption causing
the AR jet in question.  

Based on these findings, we suspect that most if not all active region jets originate from minifilament eruptions, just 
as do most or all quiet-Sun and coronal-hole jets (see references and discussion in \S\ref{sec-introduction}).  Due, however, 
to the presence of copious absorbing material that often obscures our view to the low-atmospheric levels in 
active regions, and also due to the small distance over which such a minifilament might have to erupt before reconnecting 
with far-reaching field, it is necessary for the erupting minifilaments to be positioned appropriately along the observation
line-of-sight for us to be able to see them outside of the obscuring absorbing material.  This would account for the low rate
of reporting of erupting minifilaments as the cause of jets in active regions.  This could also explain why some energetic 
active regions jets appear to develop rapidly and from very low in the corona, leading to designations as ``violent jets" 
in \citet{sterling.et17}, and ``coronal geysers" in \citet{paraschiv.et19}.

\comment{

Another factor with jets in active regions, compared to those in quiet Sun and coronal holes, is that in active regions the magnetic
field is both stronger and more rapidly evolves.  This likely leads to more rapid jet production, and situations where nearby eruptions are
very likely to contribute to the triggering of jets, just like has been observed in large-scale eruptions 
\citep[e.g.,][]{torok.et11,schrijver.et11}, perhaps via a process sometimes called {\it lid removal} (\citeauthor{sterling.et14}~\citeyear{sterling.et14}, \citeauthor{panesar.et15}~\citeyear{panesar.et15}, \citeauthor{joshi.et20}~\citeyear{joshi.et20};
also see \citeauthor{joshi.et21}~\citeyear{joshi.et21}).  
Jets in both the south and the northwest locations show evidence
of being triggered in part by eruptions on different neutral lines.  For the south-location's first main jet (event~4), the eruption that may have
assisted triggering of that main jet produced a comparatively small jet
in the south location but on a different neutral line.  For the two northwest-location main jets (events~5 and~8) it was the two 
south-region main jets (events~4 and~7) that appear to have contributed to the triggering of the onset of both of those northwest events.

} 

\comment{

Figures\,\ref{hide_zu4} and\,\ref{hide_zu4a}  show that the flux polarities in our line-of-sight magnetograms largely reproduce 
those that appear in the disambiguated vector radial-field component vector magnetograms.  Specifically, for the south region, 
the line-of-sight magnetograms agree well with the fluxes shown in the vector magnetogram.  For the northwest region, our line-of-sight
magnetograms may underestimate the amount of negative flux; despite this however, the northwest region in the vector magnetogram
also shows mixed polarity, and so our hypothesis that cancelation among mixed-polarity fluxes occurs and causes minifilaments 
to erupt and create jets remains in tact, although we are not able to present definite proof of this.

} 

There have been a few other studies of jet onset that utilized HMI vector magnetograms.  \citet{guo.et13}  studied a region about 200$''$
east and 400$''$ south of disk center, in their study of a jet-producing active region. They also used both vector and line-of-sight HMI 
magentograms in the investigation.  From this, they successfully inferred that the jets ``were associated with the quasi-separatrix layers
deduced from the magnetic extrapolations."  They also concluded that ``the magnetic reconnection occurs periodically, in the 
current layer created between the emerging bipoles and the large-scale active region field," apparently producing the jets.  Their 
included movie (their movie 3) of the line-of-sight field, however, shows clearly that the positive polarity of that emerging bipole 
runs into pre-exisiting negative field (from about 3:00 UT on 2010 September~17 in their movie 3), and this corresponds in location and
in time to the jets that they present in their Figure\,1.  Therefore, these results appear to be consistent with magnetic flux cancelation 
being the underlying cause of the jets, where one polarity of the emerging field undergoes cancelation with surrounding pre-existing opposite-polarity
field \citep[e.g.,][]{shen.et12,sterling.et16b,panesar.et18a}.  In our experience, jets are sometimes seen to occur near locations of flux emergence
\citep[e.g.,][]{sterling.et16b,panesar.et18a}, but also often at cancelation sites in the absence of emergence \citep[e.g.,][]{adams.et14}.  
Moreover, in 
the case
of emergence, it is typical that the jets occur at the site where one emerging leg of the emerging bipole cancels with surrounding 
pre-existing opposite-polarity field. Based on these studies we conclude that flux cancelation is likely the primary mechanism essential 
for jetting.

\citet{paraschiv.et20} also examined active region jets using vector magnetograms.  They studied an active region nearly
800$''$~east and 200$''$~north of disk center, and inspected  disambiguated vector magnetic fields to look at the magnetic source
of ten jets.  They concluded that 4 of the ten jets resulted from flux cancelation, and argued that the remaining six were ``clearly not"
due to flux cancelation and ``were more likely due to flux emergence."  While this is a novel approach to examining the question of
the magnetic source of jets, one must also consider the detectability limits of the observations for magnetic features near the limb
that are substantially foreshorted.  \citet{paraschiv.et20} found the noise limit for the transverse fields to be $\sim$130\,G, and this
could exceed the strength of the fields undergoing cancelation to make the jets.   We can examine our own case here for an 
estimate of how strong the canceling fields might be.  In our south region, the cancelation is occurring with the minority-polarity 
negative flux elements contained in the yellow box in Figure\,\ref{hide_zu4}(c).  There are two obvious negative-polarity (black)
patches inside that box; at the time of that magnetogram, we measure the line-of-sight average field strength of the northern black 
patch to be $\sim -40$\,G, and the southern black patch is $\sim -20$\,G\@.  At these times the fluxes are somewhat weakened 
compared to a few minutes earlier, when they were further separated from the large white (positive-polarity) patch in the yellow
box; for example, we measured the field strengths at  00:42\,UT (see movie accompanying Fig.\,\ref{hide_zu4}), finding the average 
strengths of the two negative-polarity patches to be respectively  $\sim-70$\,G and $\sim -50$\,G\@.  Due to the foreshortening  
at the  $-430'' \times 310''$ location, the vertical component of these fields would be $\sim$20\% larger
than these values.  Even with this, however, these patches would have strengths below the noise level of the $\sim$130\,G noise
level of highly foreshortened vertical fields.  In other words, if the jets observed by  \citet{paraschiv.et20} had flux patches of
similar strengths to those we observe undergoing cancelation and making jets in our case, those fields in the 
\citet{paraschiv.et20} case would be near or below the noise level due to the high foreshortening (meaning that a large component
of the disambiguated field in their case would be tangential).   In the case of non-active-region jets, the fields that undergo cancelation. 
Therefore, extreme caution should be exercised before concluding that a particular mechanism is or is not driving jets when
comparatively low strength magnetic patches cannot be reliably determined, such as from HMI vector magnetograms of regions
far from disk center.

\citet{joshi.r.et20b} examine an AR jet using vector magnetograms and AIA image, and \iris\ spectra.  They find that the 
vector magnetograms and line-of-sight magnetograms are consistent in that they both show that the jet originates from the 
site of magnetic cancelation, which is consistent with our view. (Although the question of how their proposed scenario 
for production of the jet from that cancelation compares with our scenario for jet production, and how those proposed 
scenarios compare with the observations, would required a more detailed investigation that is beyond the scope of this paper.)

\citet{schmieder.et22} also examine jets in terms of vector magnetic fields.  But those jets are observed in UV with \iris\@.
And while some of these jets may operate in a manner similar to the coronal jets that are the focus of our discussion
\citep[e.g.,][]{panesar.et22},  other jets discussed by \citet{schmieder.et22}, the so-called ``nanoflares'' \citep[cf.\,][]{antolin.et21},
are perhaps created by a different mechanism.  For this reason we do not consider the  \citet{schmieder.et22} in detail here.

Our two main jets in the south region (events~4 and~7) were both preceded by precursor eruptions (events~1---3, and event~6), 
which seemed to originate from eruptions of strands of the respective main-erupting 
minifilament field.  Such erupting strands have been seen to cause active region jets before 
also. as \citet{sterling.et17} reported that they saw several active region jets resulting from the eruption of thin minifilament ``strands"
of width $\ltsim$2$''$.  In our cases here we observe erupting minifilament strands that are perhaps a little thinner than those previously reported ones, being of
width $\sim$1---2$''$.  \citet{schmieder.et13} observe a pattern of ``dark and light strands" propagating along an active region 
jet, and other workers have pointed out the multi-stranded nature of jets \citep[e.g.][]{mulay.et16}, and those may be similar to the features 
that we report here.

In summary: our findings for this region show that when we can observe AR jets at their origin (such as those in the south location 
in the current study), they follow the minifilament-eruption 
picture for jets, where the minifilaments are likely formed and triggered to erupt via magnetic flux cancelation.  Those minifilaments 
erupt and undergo external reconnection after traveling only a short distance from their eruption location, at which time the 
minifilament character (appearing as a filament-like feature rooted on two ends in the photosphere) becomes lost, and the cool 
minifilament material then flows outward along the jet's spire.  Only for those AR jets that are positioned fortuitously along our line-of-sight, 
however, can we see down to their origin location.  For other AR jets, such as the northwest-location jets in the current study, the 
origin locations are largely obscured by foreground 
absorbing low-solar-atmospheric matter, that resides above the still-lower altitude at which the pre-eruption minifilaments 
form.  In such cases, we only see the AR jets after the material has lost its minifilament character and is flowing out along the jet's spire.  

Because low-altitude EUV-absorbing material appears to be common in ARs, for AR jets for which the origin site is obscured
it is hard to ``prove" conclusively that the jets originate from minifilament eruptions, perhaps triggered to erupt by magnetic flux 
cancelation.  Our work here, however, along with the evidence for this process occurring in non-AR jets (\S\ref{sec-introduction}),
leads us to propose that virtually all coronal jets work in basically this fashion.

\vspace{1cm}

\comment{

} 

\begin{acknowledgments}
The authors thank N. Nitta, who alerted us to the jetting activity in NOAA~AR\,12824.  We thank an anonymous referee who, among 
other helpful points, alerted us to the importance of considering vector magnetograms in our work and of discussing other works 
studying jets with vector magnetograms.  A.C.S., 
R.L.M. and N.K.P. received funding from the Heliophysics Division of NASA's Science 
Mission Directorate through the Heliophysics Supporting Research (HSR, grant No.~20-HSR20\_2-0124) Program, 
and the Heliophysics Guest Investigators program.  N.K.P. received additional support through a NASA \sdo/AIA grant.   
We acknowledge the use of AIA data. AIA is an instrument onboard \sdo, a mission of
NASA's Living With a Star program.
\end{acknowledgments}

\bibliography{refs_hide}

\clearpage


\begin{deluxetable}{lllllll@{\hskip 1.5cm}l}
\tabletypesize{\scriptsize}
\tablecaption{Jet-like Event Sources in AR~12824 \label{tab:table1}}
\tablehead{
\colhead{Event}& \colhead{Time (UT)\tablenotemark{a}}& \colhead{Event Type} & \colhead{\goes\ Level}&  \colhead{Location\tablenotemark{b}} &  \colhead{Notes}
}
\startdata
1 & 00:45 &  Precursor & B1 & South & Small eruption from erupting minifilament strand. \\
2 & 01:30 &  Precursor & B1 & South & Small eruption from erupting minifilament strand. \\
3 & 02:10 &  Precursor & $\approx$B2 & South & Small eruption from erupting minifilament strand. \\
4 & 02:55 &  Main         & C6 & South & Eruption of minifilament. \\
5 & 03:35 &  Main         & $\approx$B4 & Northwest & Indeterminant. \\
6 & 05:40 &  Precursor & $\approx$B2 & South & Small eruption from erupting minifilament strand. \\
7 & 06:15 &  Main         & C6 & South & Eruption of minifilament. \\
8 & 06:55 &  Main         & C1 & Northwest & Indeterminant. \\
\hline
\hline
\enddata
\tablenotetext{a}{UT time of event peak on 2021 May 22.}
\tablenotetext{b}{Event location in AR~12824.}
\end{deluxetable}
\clearpage


\begin{figure}
\centering
\includegraphics[width=\textwidth,angle=0]{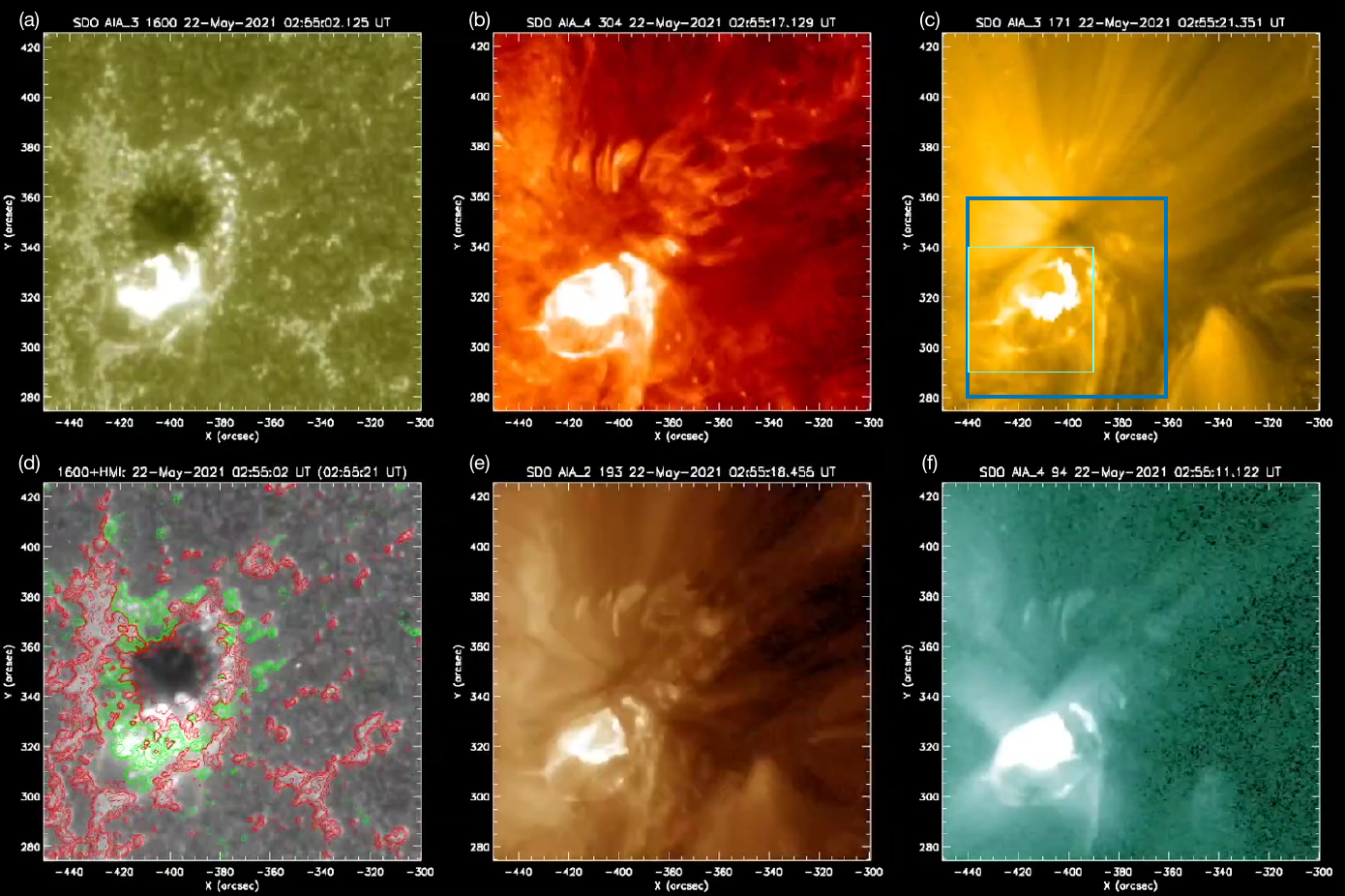}
\caption{\sdo/AIA and HMI observations of NOAA AR~12824 at the time of a jet-producing eruption on the south 
side of the sunspot.  All six panels are at approximately the same time, showing AIA images from channels (a) 1600,
(b) 304, (c) 193, (d) 1600, (e) 193, and (f) 94\,\AA\@.  Times of the images appear above each respective panel.  
Panel~(d) has overlaid onto the 1600\,\AA\ image contours from an HMI magnetogram of the time given 
in parentheses above the panel, with contour levels the same as those of Fig.\,\ref{hide_zu4}(b).  The eruption is 
obvious in all panels, on the south side of the sunspot visible in the 1600\,\AA\ images. Panels (b), (c), (e) and (f) show
a jet being expelled from the west side of the brightening with the spire directed toward the south. North is upward and west is 
to the right in these and in all other solar images in this paper.  The blue box in (c) shows the field of view (FOV) of the zoomed-in 
figures below, and the turquoise box in (c) shows the area over which the light curve
in Fig.\,\ref{hide_zu3}(b) is created.  The accompanying animation shows the evolution of these
panels, covering 2021 May~22 0---8\,UT, with time cadence of 96\,s.  The total duration of the animation is 10\,s. 
(The animation is the same for Figs.~1 and~2.)}
\label{hide_zu1}  
\end{figure}
\clearpage

\begin{figure}
\centering
\includegraphics[width=\textwidth,angle=0]{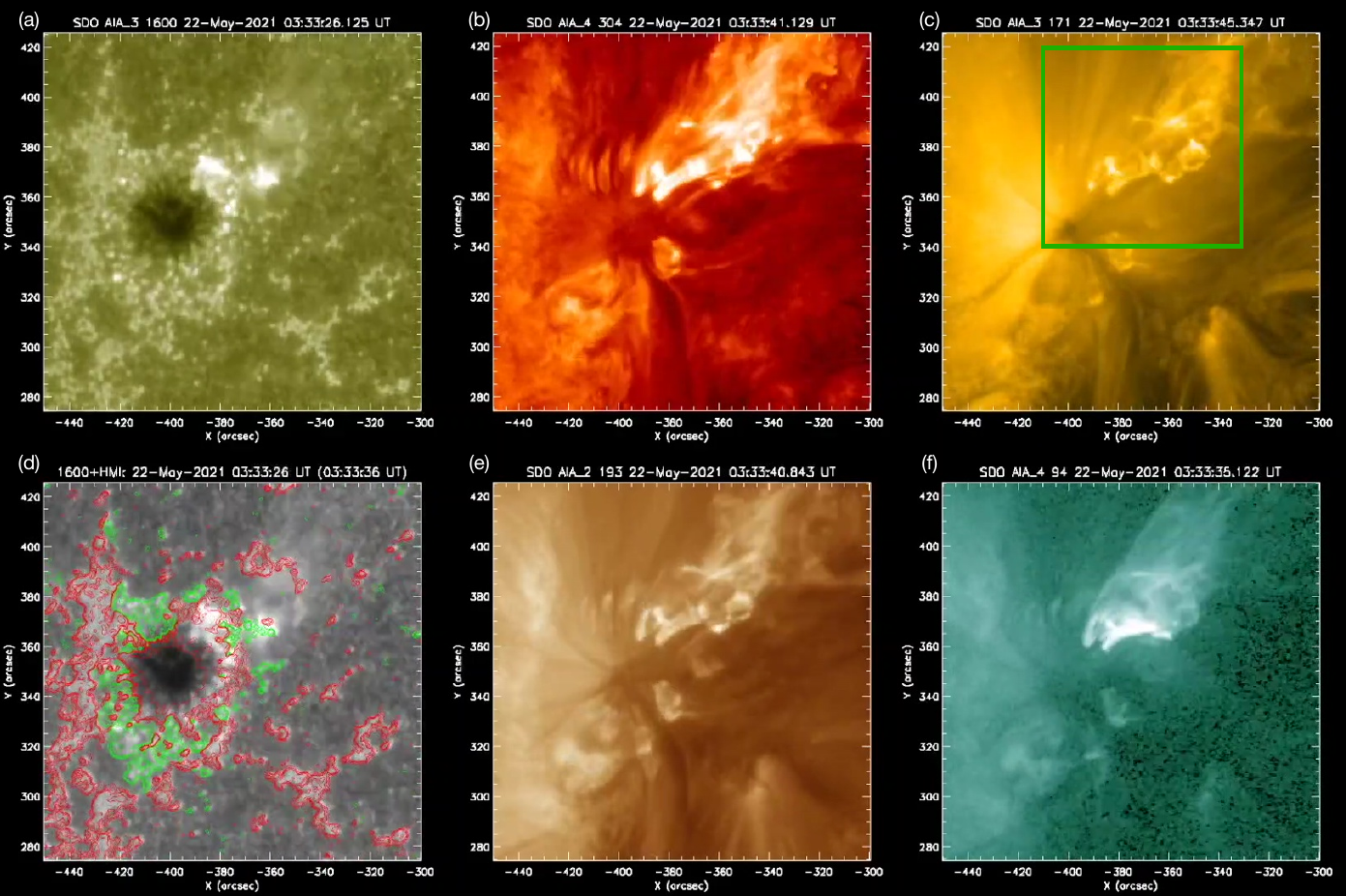}
\caption{\sdo/AIA and HMI observations of NOAA AR~12824 at the time of a jet-producing eruption on the northwest 
side of the sunspot.  The layout of this figure is the same as that of Fig.\,1, but for a time when a jet-producing eruption
is occurring on the northwest side of the spot visible in the 1600\,\AA\ images.  The green box in (c) shows both the
FOV of the zoomed-in images in the figures below, and also the area 
over which the light curve in Fig.\,\ref{hide_zu3}(a) is created. The accompanying animation shows the evolution of these
panels, covering 2021 May~22 0---8\,UT, with time cadence of 96\,s.  The total duration of the animation is 10\,s. 
(The animation is the same for Figs.~1 and~2.)}
\label{hide_zu2}  
\end{figure}
\clearpage

\begin{figure}
\centering
\includegraphics[width=12cm,angle=0]{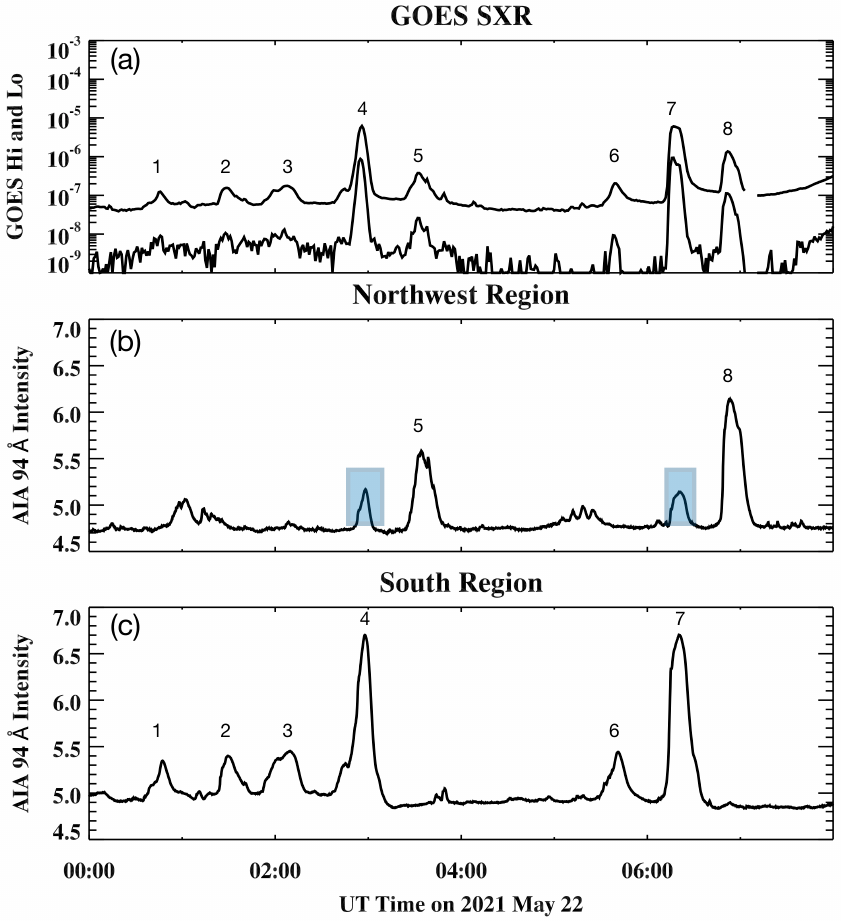}
\caption{Intensity changes with time from the events.  (a) \goes\ high (0.5---4.0\,\AA) (lower profile) and low (1---8\,\AA) (upper profile) 
soft X-ray full-Sun flux 
changes over the duration of observation period. The units are the standard W\,m$^{-2}$, with 1---8\,\AA\ $\log_{10}$ 
values of $-7$, $-6$, and 
$-5$ representing \goes\ B, C, and M classes, respectively.  (b) \sdo/AIA~94\,\AA-channel lightcurve of the sub-location of AR~12824 
on  the northwest side of the sunspot, calculated over the green boxed location in Fig.\,\ref{hide_zu2}(c).  (c) As in (b), but the 
\sdo/AIA~94\,\AA-channel lightcurve from the location south of the sunspot, calculated over the turquoise-color boxed location 
in Fig.\,\ref{hide_zu1}(c).  In (b), the blue-shaded region show peaks that are 
not from emission from the eruptions in the green box; these are false intensity peaks resulting from scattered light from the 
bright emission in the south region occurring at the same respective times.  These are the result of scattered light from, 
respectively, events~4 and~7 entering into the FOV of the 94\,\AA\ intensity box of the north location used to make the plot in (b).   The numbers on the peaks correspond to the event numbers
in Table~\ref{tab:table1}.}
\label{hide_zu3}  
\end{figure}
\clearpage

\begin{figure}
\centering
\includegraphics[width=\textwidth,angle=0]{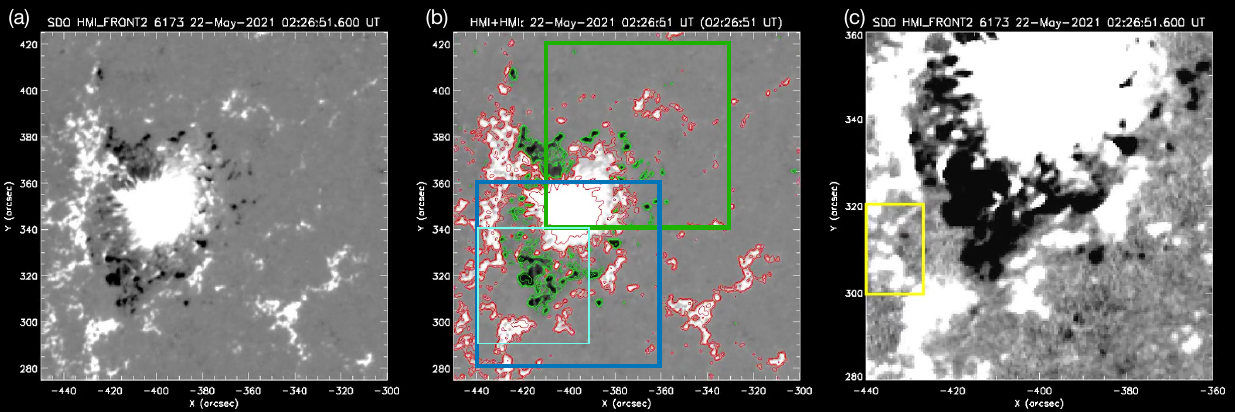}
\caption{\sdo/HMI magnetogram of the region shown in Figs.\,\ref{hide_zu1} and~\ref{hide_zu2}, with the same FOV as
in those figures, at the time of the images in Fig.\,\ref{hide_zu1}.  Panel~(a) shows the magnetogram with white (black) 
representing positive (negative) polarity, 
with the saturation level set at $\pm$300\,G\@.  Panel~(b) shows the same magnetogram, overlaid with contours of
levels $\pm$50, 100, and 500\,G; these are the same contour levels used in the overlays in 
Figs.\,\ref{hide_zu1}(d) and~\ref{hide_zu2}(d). The blue and turquoise boxes are as in Fig.~\ref{hide_zu1}, and the green 
box is as in Fig.~\ref{hide_zu2}.  Panel~(c) shows a close up of the south location, with the entire frame having the FOV of 
the blue box in (b).  Here, the saturation is set to $\pm$50\,G to highlight weaker fluxes.  The yellow box
shows the area over which the negative magnetic flux is plotted in Fig.~\ref{hide_zu7}.  The accompanying animation shows 
the evolution of these
panels, covering 2021 May~22 0---8\,UT, with time cadence of 90\,s.  The total duration of the animation is 11\,s.}
\label{hide_zu4}  
\end{figure}
\clearpage

\begin{figure}
\centering
\includegraphics[width=10cm,angle=0]{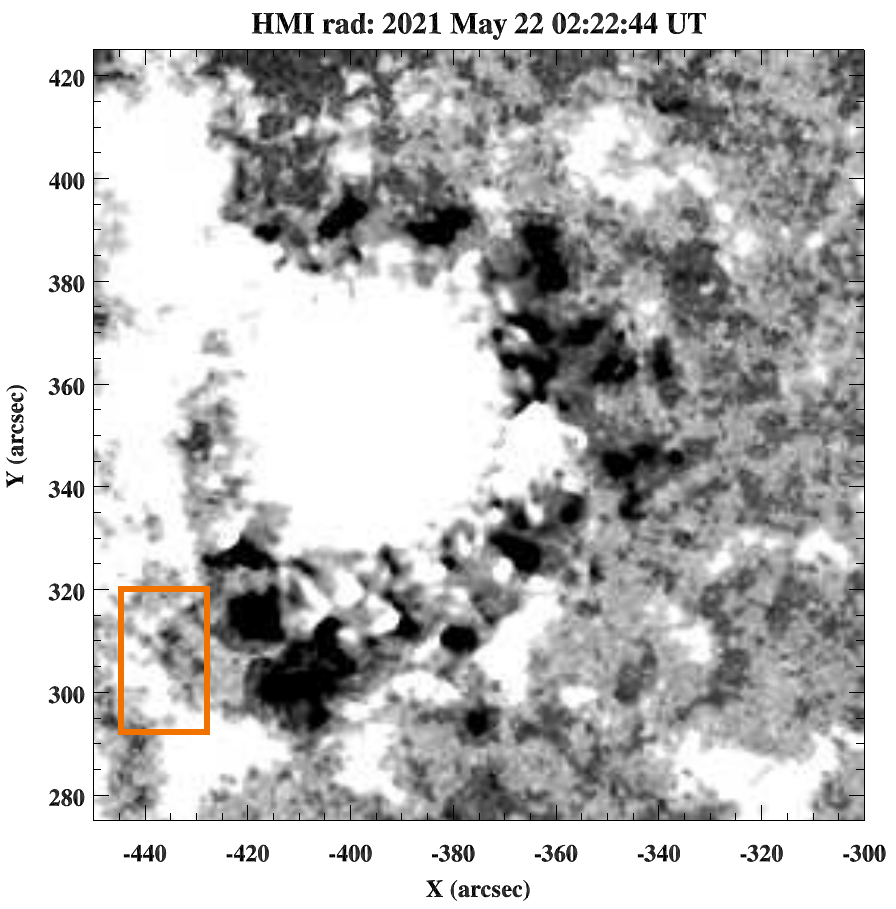}
\caption{Radial component of an \sdo/HMI vector magnetogram of the region shown in Fig.\,\ref{hide_zu4}, at about the
same time.  This shows the radial field constructed from the three vector components.  The saturation is
set at $\pm$100\,G to match approximately features visible in Fig.\,\ref{hide_zu4}.   In the southern region, 
the orange box corresponds to the yellow box in  Fig.\,\ref{hide_zu4}(c), and presence of these same comparatively weak 
fluxes shows that the projection effects in the line-of-sight magnetograms of Fig.\,\ref{hide_zu4} are small
in the southern region.  In the northwest region however, there is more widely distributed negative flux than is visible in
the line-of-sight flux of Fig.\,\ref{hide_zu4}, and so we use additional caution in drawing conclusions from the northwest line-of-sight 
fluxes.}
\label{hide_zu4a}  
\end{figure}
\clearpage

\begin{figure}
\centering
\includegraphics[width=\textwidth,angle=0]{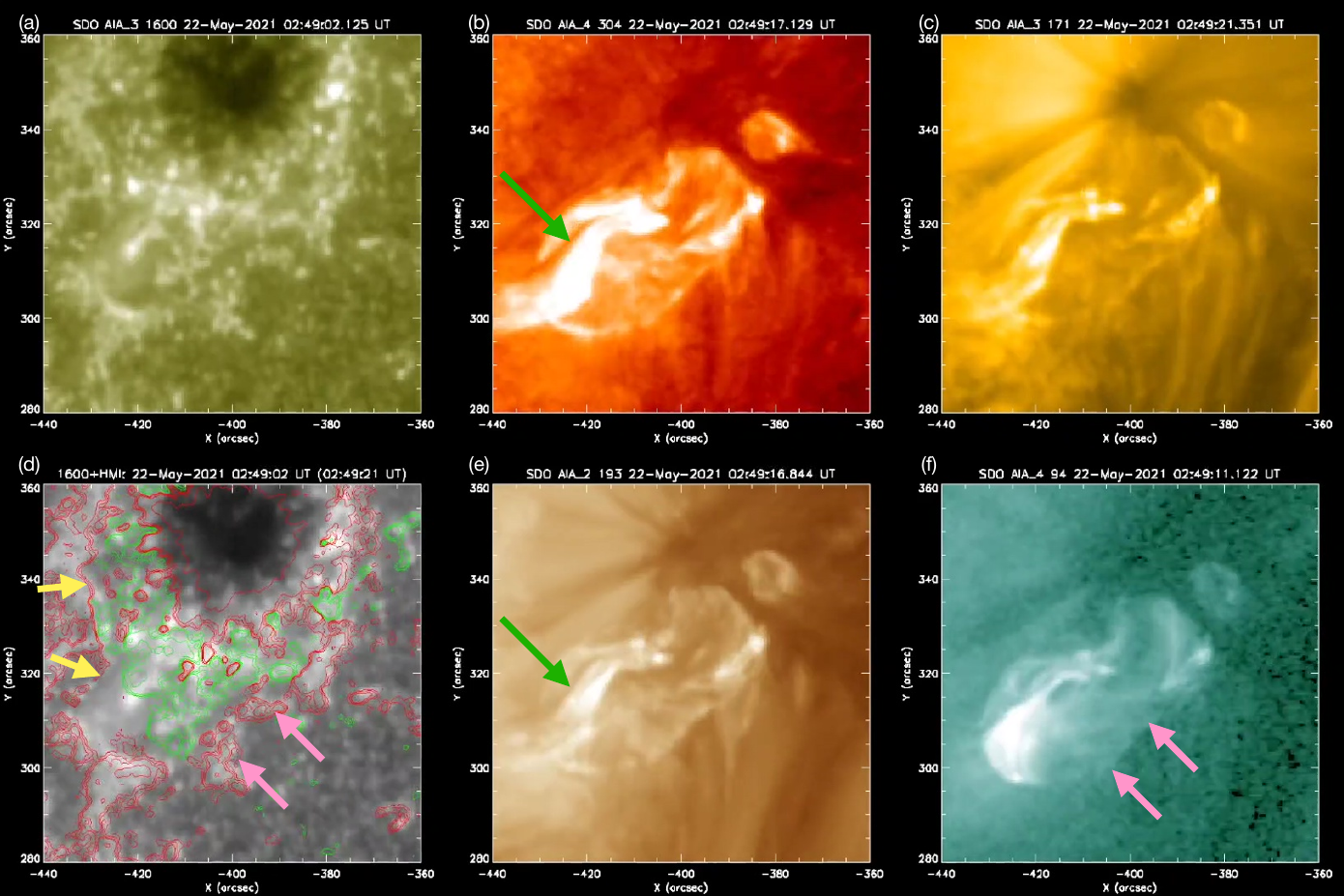}
\caption{Close up of the southern location of the active region, with the FOV of the blue box in Fig.\,\ref{hide_zu1}(c),
at the time of the first large jet, event~4 in Table~\ref{tab:table1}.  The layout of the panels and AIA wavelengths are the same as in 
Figs.\,\ref{hide_zu1} and~\ref{hide_zu2}, with the exception that the HMI contours in (d) include $pm$25\,G, in addition
to $\pm$50, 100, and 500\,G\@.   Green arrows in (b) and (e) point to an erupting minifilament,
which is also visible in (c).  Yellow arrows in~(d) point to a magnetic neutral line from which part of the
erupting minifilament originated.  Magenta arrows in~(d) point to the west edge of a bright circular ribbon, that if
also visible in (b), (c), (e), and (f), with the magenta arrows placed at the same location in (f) and (d).  The accompanying 
animation shows the evolution of these panels, covering 2021 May~22 0---8\,UT, with time cadence of 48\,s, and 
the total duration of the animation is 20\,s.}
\label{hide_zu5}  
\end{figure}
\clearpage

\begin{figure}
\centering
\includegraphics[width=14cm,angle=0]{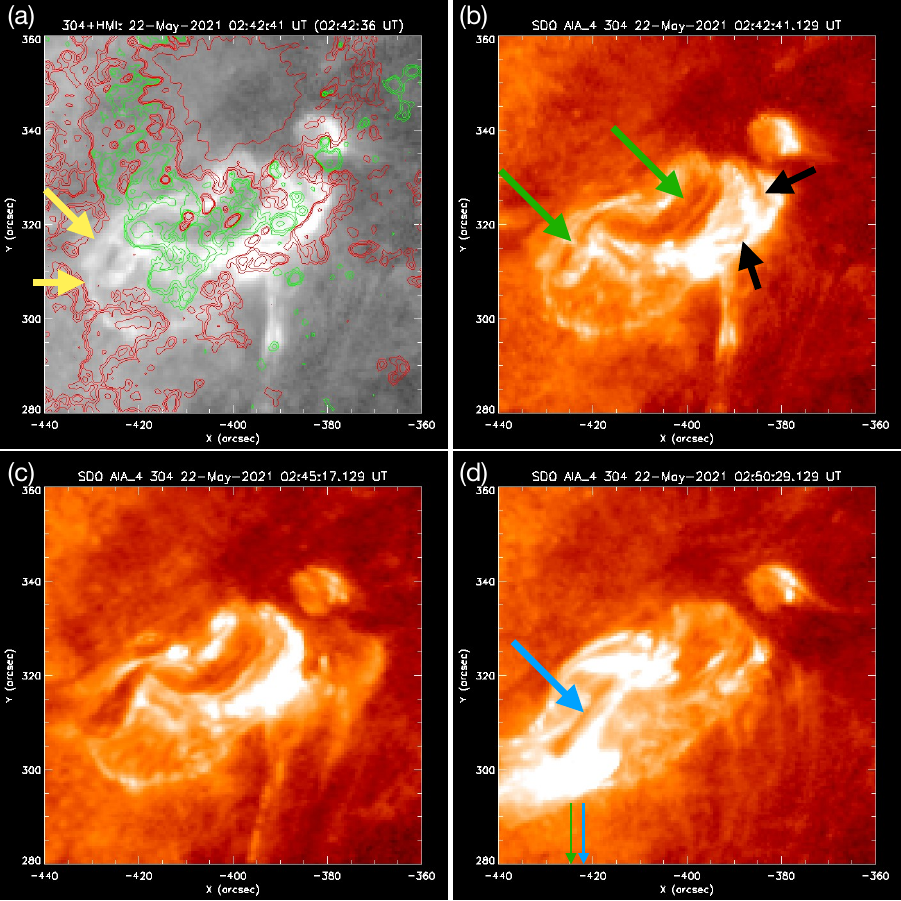}
\caption{AIA~304\,\AA\ channel images with the same FOV as the blue box in Fig.\,\ref{hide_zu1}(c), and the panels in 
Fig.\,\ref{hide_zu5}.  Panels~(a) and~(b) are at the same time, but (a) has an HMI magnetogram, with contours the
same as those in Fig.\,\ref{hide_zu5}, i.e.\ $\pm$25, 50, 100, and 500\,G\@.  Green arrows in (b) show the minifilament
that is starting the erupt, near the time that the filament strand that the lower arrow points to has just started to become
apparent. In the same panel, the black arrows show the location of a nearby jet-producing eruption. In (c) this minifilament has continued its evolution, and in (d) it is starting to erupt outward toward the southeast,
along the spire of the AR jet, as that spire is still sweeping toward the west (see the accompanying animation).  
In (d) the upper (thicker) blue arrow points to the same strand as the lower green arrow in (b), and between
these two frames the minifilament has moved in the plane of the image from east to west; as is
apparent in the animation,
it is thus over about this distance that it travels from its first appearance until the time when it takes on the appearance 
as part of the spire and loses its ``filament" character.  The thin-green and thin-blue arrows at the bottom of (d) project the heads of respectively the panel-(b) lower-green
arrow and the panel-(d) upper-blue arrow onto the abscissa axis, showing that the horizontal displacement from the minifilament 
pre-eruption location where the erupting minifilament has lost its filament character is only about $3''$.  The accompanying 
animation shows the evolution of these panels, covering 2021 May~22 0---8\,UT, with time cadence of 12\,s, and 
the total duration of the animation is 40\,s.}
\label{hide_zu6}  
\end{figure}
\clearpage

\begin{figure}
\centering
\includegraphics[width=14cm,angle=0]{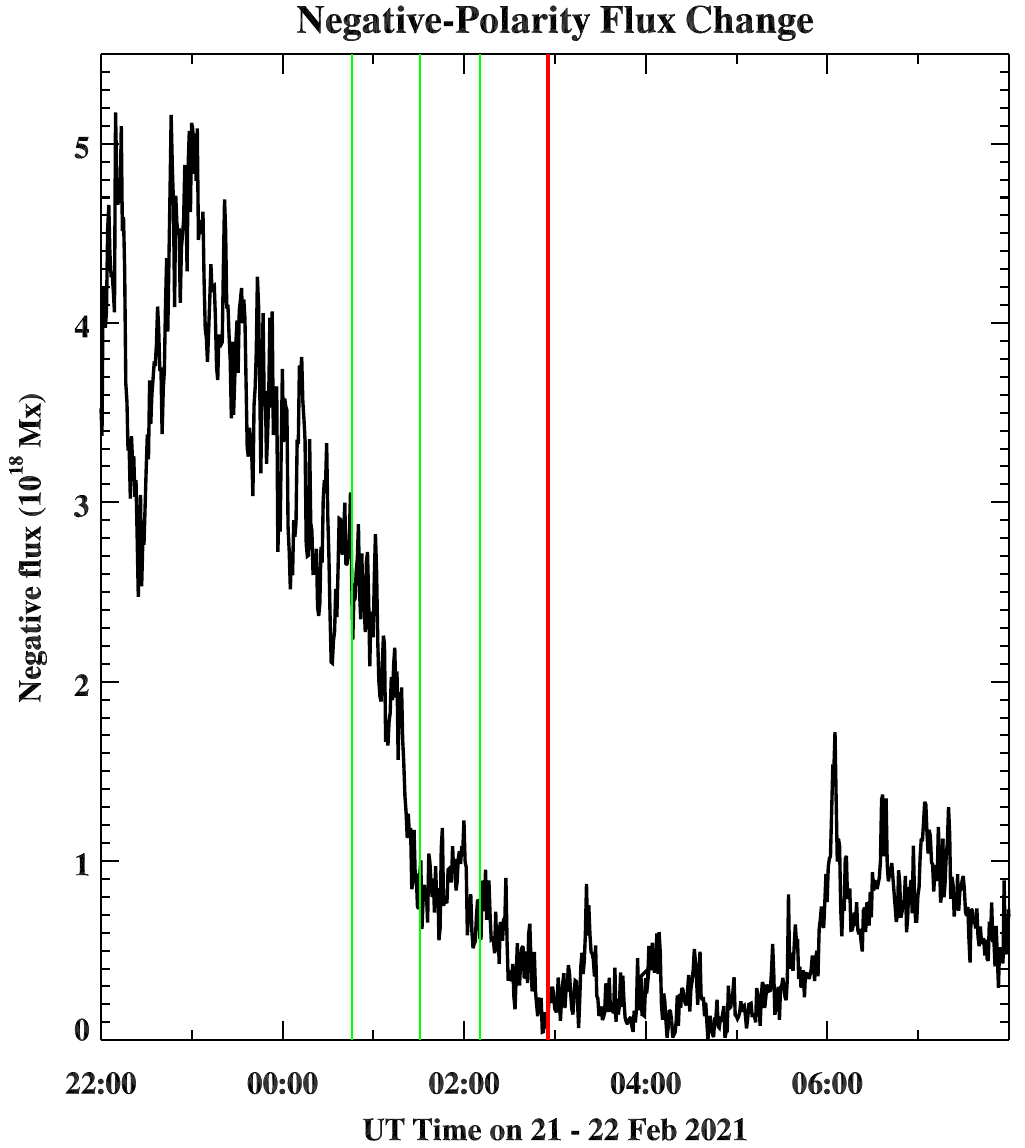}
\caption{Change in negative-polarity flux over the subregion of the south location of the active region for the first main event, where we sum over the
negative flux of strength $\gtsim 10$\,G contained in the yellow box in Fig.~\ref{hide_zu4}(c), at each timestep. We avoid negative fluxes 
between 0 and $10$\,G to filter out noise, and we have compensated for the $\cos \theta$ foreshortening effect resulting from the off-center 
location of the region.
Vertical lines denote the times of the eruptions that originate from
this location, and listed as events~1---4 in Table~\ref{tab:table1}, where the green lines are for precursor events and the red line
is for the main eruption.  The south region's second main event is also suspected to have been initiated by magnetic flux cancelation, but 
the canceling locality is outside of the field of view covered by this plot, i.e.\ outside of the yellow box in  Fig.~\ref{hide_zu4}(c) (see text), and hence its onset
time in not marked in this plot.}
\label{hide_zu7}  
\end{figure}
\clearpage

\begin{figure}
\centering
\includegraphics[width=14cm,angle=0]{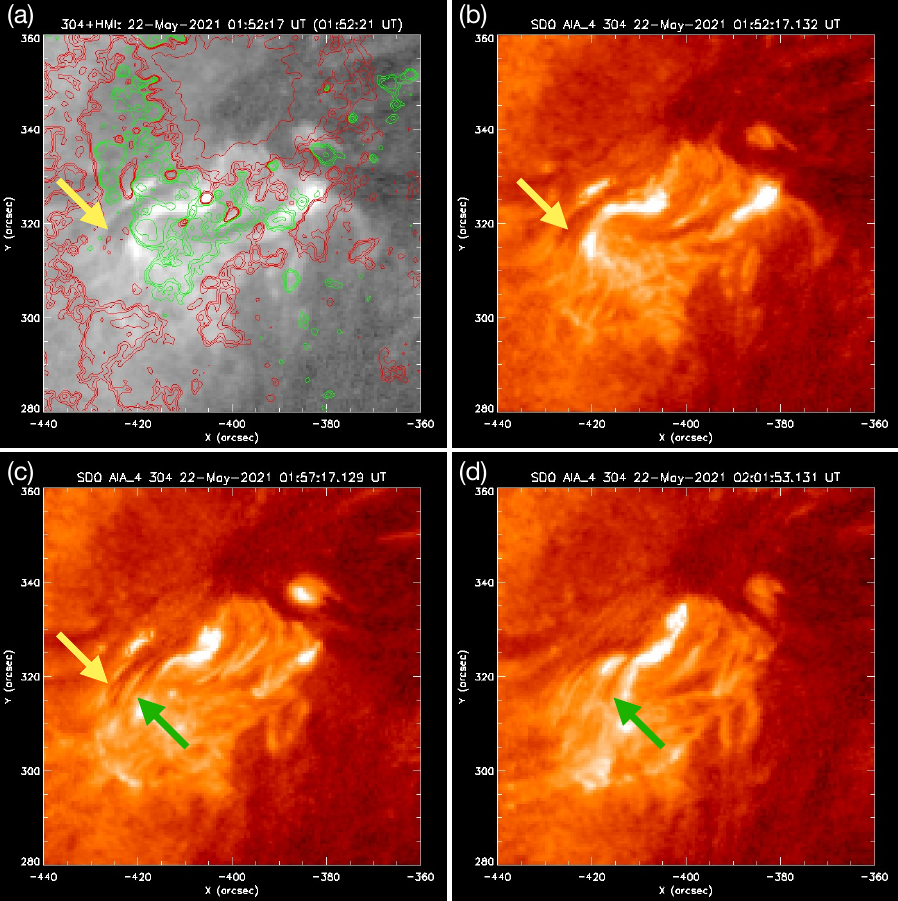}
\caption{Evolution of event number 3 in Table~\ref{tab:table1}, the third precursor event prior to the first main eruption.  The layout is the same
as in Fig.~\ref{hide_zu6}.  The yellow arrows are all in the same location in (a), (b), and (c), and show that part of a (mini)filament
remains in about the same location during the event, at a location near a magnetic neutral line in (a). The green arrows 
point to a portion of that minifilament that appears to peal off from the main minifilament (c) and partially eject outward (d).}
\label{hide_zu8}  
\end{figure}
\clearpage

\begin{figure}
\centering
\includegraphics[width=14cm,angle=0]{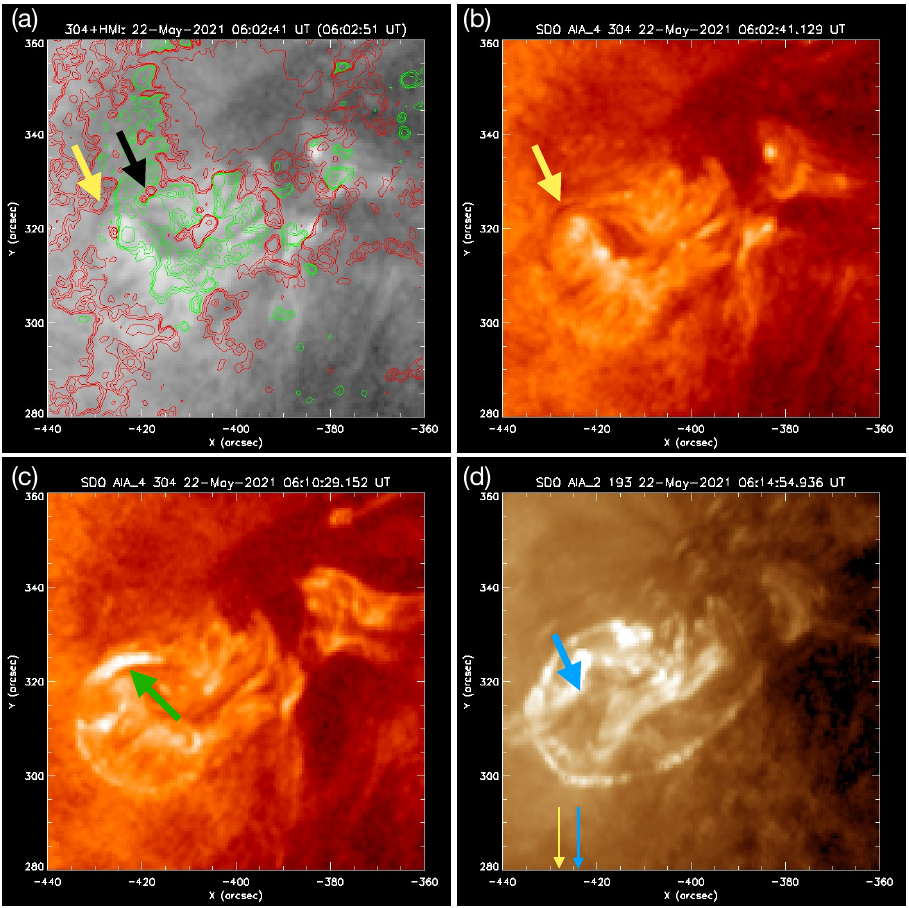}
\caption{Evolution of event number 7 in Table~\ref{tab:table1}, the second main eruption in the south location.  The layout is the same
as in Figs.~\ref{hide_zu6} and~\ref{hide_zu8}.  The yellow arrows are all in the same location in (a) and (b), pointing to the 
minifilament that erupts in the making of the jet. In (c) the green arrow shows the minifilament starting to erupt, with a brightening starting
at the location where it had been in (a) and (b), which would be due to flare-like reconnection below the erupting minifilament. 
In (d), the minifilament, pointed to by the top blue arrow, has moved further, and has now started to flow out along extended field 
lines as a jet spire.  The thin-yellow and thin-blue arrows at the bottom of (d) project the heads of respectively the panel-(b) yellow
arrow and the panel-(d) top-blue arrow onto the abscissa axis, showing that the horizontal displacement from the minifilament 
pre-eruption location where the erupting minifilament has lost its filament character is only about $4''$. In (a), there are two 
locations that are candidates for flux cancelation that could have destabilized the minifilament's magnetic structure and caused
its eruption: the black arrow points to two small positive-flux patches, the left-most of which disappears and the right-most of which 
shrinks during the time of the eruption; and the neutral line (red-green-contour boundary) just west of the yellow arrow is dynamic over
time, with the two polarities flowing into each other over time.  These magnetic dynamics can be seen 
by comparing with the animation accompanying Fig.\,\ref{hide_zu4}(c).  A 12-s-cadence movie of 
304\,\AA\ images is available in the animation accompanying Fig.\,\ref{hide_zu6}.}
\label{hide_zu9}  
\end{figure}
\clearpage

\begin{figure}
\centering
\includegraphics[width=\textwidth,angle=0]{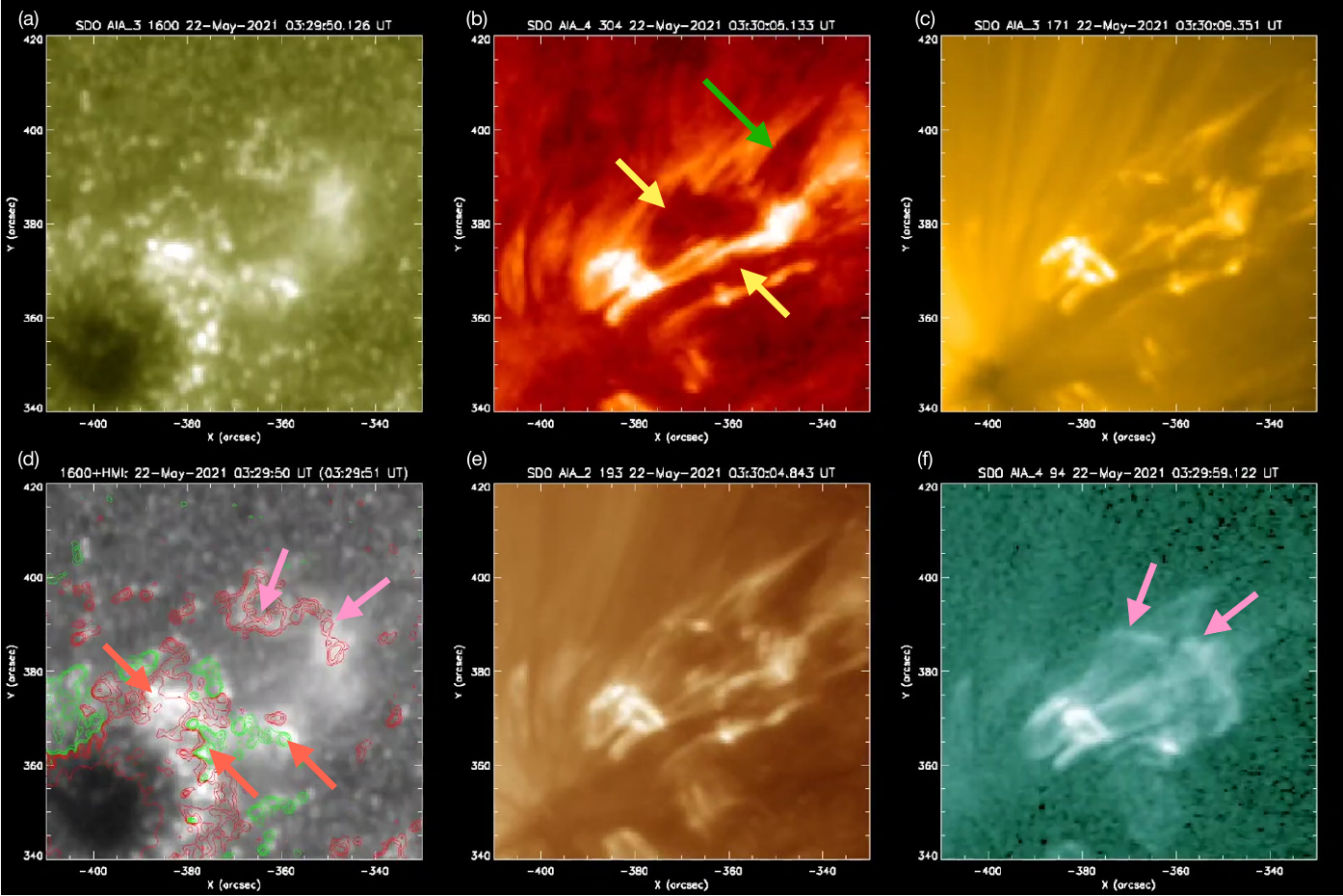}
\caption{Evolution of event number 5 in Table~\ref{tab:table1}, the first main eruption in the northwest location.  The layout is the same
as in Figs.~\ref{hide_zu1}, \ref{hide_zu2}. and~\ref{hide_zu5}.  In (b), the green arrow points to cool material ejecting out onto the 
spire of the jet.  In this case we do not see where the eruption originates, due to copious obscuring material, with the yellow arrows
in (b) pointing to two examples of obscuration.  Based on brightening in
the accompanying animation, likely source locations are mixed-polarity areas pointed to by the orange arrows in (d).  In this case
the ejected material does not have the appearance of an erupting minifilament.  But even if the material pointed to by the green arrow 
in (b) was originally a minifilament at one of the location of one or more of the orange arrows in (d), and if that minifilament shared the
same properties as the minifilaments that erupted to form the jets in the south location, then that minifilament would have lost its ``filament"
character after a distance of about $5''$ from the orange-arrow locations; but all such locations would be obscured from our view due
to the absorbing material along the line of sight from Earth.   Thus, we would be unable to recognize that the jet originated from an 
erupting minifilament in this case, despite it having done so.  Magenta arrows 
in (d) point to brightenings that, together with the orange arrows, forms a partial-circular ribbon structure, similar
to the circular structures of Fig.~\ref{hide_zu5}.  The accompanying 
animation shows the evolution of these panels, covering 2021 May~22 0---8\,UT, with time cadence of 48\,s, and 
the total duration of the animation is 20\,s.}
\label{hide_zu10}  
\end{figure}
\clearpage

\begin{figure}
\centering
\includegraphics[width=\textwidth,angle=0]{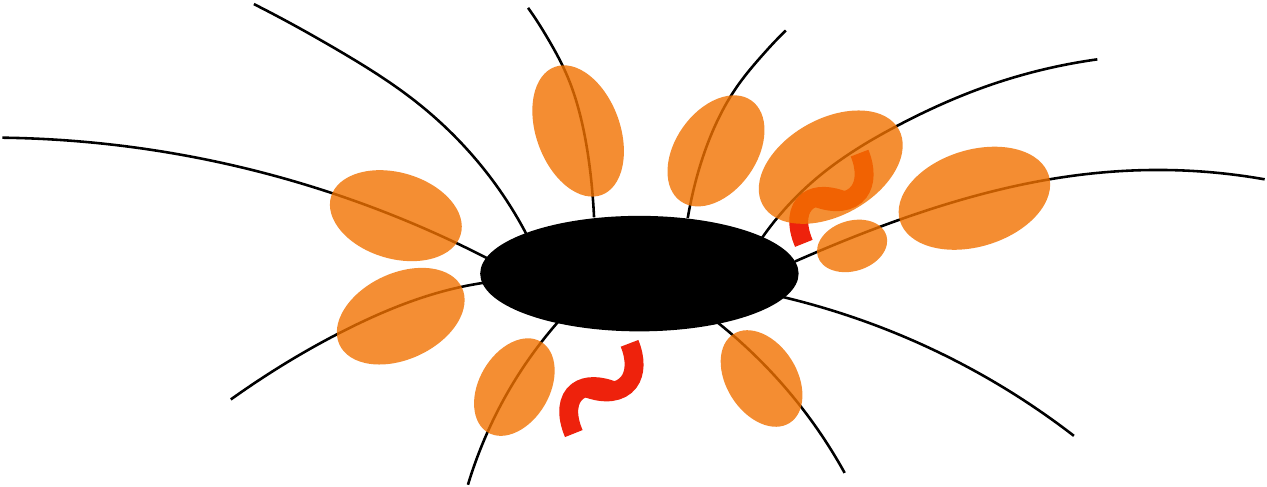}
\caption{
Schematic of the observed region, showing the sunspot and its neighborhood with the same orientation as 
in the solar images in the figures, with north up and west to the right.  The black oval is the sunspot umbra 
(the penumbra is not pictured).  Black lines are low-lying magnetic field lines fanning out from the edge of 
the umbra at a low angle to the solar surface; other field lines, such as ones that are rooted farther inside 
the umbra and extend more vertically, are not pictured.  Orange ovals are elevated patches of 
chromospheric-temperature plasma.  These are suspended in the low-angle field threading them and are 
largely opaque to AIA EUV wavelengths.  Red sigmoid-shaped worms are magnetic flux ropes/minifilaments 
that are poised to erupt and produce jets.  These minifilaments apparently form low in the atmosphere, 
below the heights of the elevated opaque plasma patches.  Due to the viewing angle, an about-to-erupt 
minifilament in the south is more likely be visible to us than one in the northwest, because in the northwest 
the pre-eruption minifilament is more likely to be obscured by elevated opaque plasma along the line of sight.}
\label{hide_zu11}  
\end{figure}
\clearpage

\end{document}